# GOLDRUSH. II. Clustering of Galaxies at $z \sim 4-6$ Revealed with the Half-Million Dropouts Over the 100 deg$^2$ Area Corresponding to 1 Gpc$^3$

Yuichi HARIKANE[1,2], Masami OUCHI[1,3], Yoshiaki ONO[1], Shun SAITO[4], Peter BEHROOZI[5], Surhud MORE[3], Kazuhiro SHIMASAKU[6,7], Jun TOSHIKAWA[1], Yen-Ting LIN[8], Masayuki AKIYAMA[9], Jean COUPON[10], Yutaka KOMIYAMA[11], Akira KONNO[1,6], Sheng-Chieh LIN[8], Satoshi MIYAZAKI[11,12], Atsushi J. NISHIZAWA[13], Takatoshi SHIBUYA[1], and John SILVERMAN[3]

[1] Institute for Cosmic Ray Research, The University of Tokyo, 5-1-5 Kashiwanoha, Kashiwa, Chiba 277-8582, Japan

[2] Department of Physics, Graduate School of Science, The University of Tokyo, 7-3-1 Hongo, Bunkyo, Tokyo, 113-0033, Japan

[3] Kavli Institute for the Physics and Mathematics of the Universe (Kavli IPMU, WPI), University of Tokyo, Kashiwa, Chiba 277-8583, Japan

[4] Max-Planck-Institut für Astrophysik, Karl-Schwarzschild-Starße 1, D-85740 Garching bei München, Germany

[5] Department of Physics, University of California, Berkeley, CA 94720

[6] Department of Astronomy, Graduate School of Science, The University of Tokyo, 7-3-1 Hongo, Bunkyo, Tokyo 113-0033, Japan

[7] Research Center for the Early Universe, Graduate School of Science, The University of Tokyo, 7-3-1 Hongo, Bunkyo, Tokyo 113-0033, Japan

[8] Institute of Astronomy and Astrophysics, Academia Sinica, 11F of AS/NTU Astronomy-Mathematics Building, No.1, Sec. 4, Roosevelt Rd, Taipei 10617, Taiwan, R.O.C.

[9] Astronomical Institute, Tohoku University, Aramaki, Aoba-ku, Sendai, 980-8578

[10] Department of Astronomy, University of Geneva, ch. d'Écogia 16, 1290 Versoix, Switzerland

[11] National Astronomical Observatory of Japan, 2-21-1 Osawa, Mitaka, Tokyo 181-8588, Japan

[12] SOKENDAI(The Graduate University for Advanced Studies), Mitaka, Tokyo, 181-8588, Japan

[13] Institute for Advanced Research, Nagoya University, Chikusaku, Nagoya 464-8602, Japan

*E-mail: hari@icrr.u-tokyo.ac.jp

## Abstract

We present clustering properties from 579,492 Lyman break galaxies (LBGs) at $z \sim 4-6$ over the 100 deg$^2$ sky (corresponding to a 1.4 Gpc$^3$ volume) identified in early data of the Hyper Suprime-Cam (HSC) Subaru strategic program survey. We derive angular correlation functions (ACFs) of the HSC LBGs with unprecedentedly high statistical accuracies at $z \sim 4-6$, and compare them with the halo occupation distribution (HOD) models. We clearly identify significant ACF excesses in $10'' < \theta < 90''$, the transition scale between 1- and 2-halo terms, suggestive of the existence of the non-linear halo bias effect. Combining the HOD models and previous clustering measurements of faint LBGs at $z \sim 4-7$, we investigate dark-matter halo mass ($M_h$) of the $z \sim 4-7$ LBGs and its correlation with various physical properties including the star-formation rate (SFR), the stellar-to-halo mass ratio (SHMR), and the dark matter accretion rate ($\dot{M}_h$) over a wide-mass range of $M_h/M_\odot = 4 \times 10^{10} - 4 \times 10^{12}$. We find that the SHMR increases from $z \sim 4$ to 7 by a factor of $\sim 4$ at $M_h \simeq 1 \times 10^{11} \, M_\odot$, while the SHMR shows no strong evolution in the similar redshift range at $M_h \simeq 1 \times 10^{12} \, M_\odot$. Interestingly, we identify a tight relation of $SFR/\dot{M}_h - M_h$ showing no significant evolution beyond 0.15 dex in this wide-mass range over $z \sim 4-7$. This weak evolution suggests that the $SFR/\dot{M}_h - M_h$ relation is a fundamental relation in high-redshift galaxy formation whose star formation activities are regulated by the dark matter mass assembly. Assuming this fundamental relation, we calculate the cosmic star formation rate densities (SFRDs) over $z = 0 - 10$ (a.k.a. Madau-Lilly plot). The cosmic SFRD evolution based on the fundamental relation agrees with the one obtained by observations, suggesting that the cosmic SFRD increase from $z \sim 10$ to $4 - 2$ (decrease from $z \sim 4 - 2$ to 0) is mainly driven by the increase of the halo abundance (the decrease of the accretion rate).

# 1 Introduction

Understanding galaxy formation and evolution is one of the important goals in modern astronomy. In the frameworks of $\Lambda$ cold dark matter ($\Lambda$CDM) structure formation models, galaxies are thought to be formed in their dark matter halos through gas cooling (Rees & Ostriker 1977). Since the efficiency of the gas cooling depends on the virial temperature and the gas number density which are related to the gravitational potential of the dark matter halo (Silk & Wyse 1993; Sutherland & Dopita 1993), the dark matter halo mass, $M_h$, is a key quantity to understand the galaxy formation physics. After their formation, galaxies evolve across cosmic time experiencing the gas accretion, feedback, and merger. These processes are also closely related to the host dark matter halo. If the baryonic gas accretes onto halos with dark matter, the average gas accretion rate is expected to be proportional to the dark matter accretion rate, $\dot{M}_h \propto M_h^{1.1}$, from N-body simulations (e.g., Fakhouri et al. 2010; Behroozi et al. 2013b). The supernovae (SN) and active galactic nuclei (AGN) feedbacks are thought to suppress star formation in low- and high-mass halos (Murray et al. 2005; Dekel et al. 2009; Kereš et al. 2009; Sugahara et al. 2017), and the mass of the host halo (and its gravitational potential) is an important factor in the efficiency of the feedback processes. In addition, both major and minor merger rates are functions of $M_h$ (e.g., Fakhouri et al. 2010). Thus, investigating the galaxy-dark matter halo connection in a wide mass range across cosmic time gives strong constraints on galaxy formation and evolution models.

There are primarily three methods widely used to statistically estimate the dark matter halo masses of galaxies (see also the satellite kinematics method; More et al. 2011, rotational curve method; Miller et al. 2014; Sofue 2016). The first is the weak lensing. In the weak lensing analysis, one measures the tangential shear of background galaxies by stacking, and calculates the halo mass needed to reproduce the observed weak lensing signals. The weak lensing analysis allows us to estimate the dark matter halo mass without any strong assumptions, because we can see the effect of the gravity directly. However, the weak lensing analysis cannot be applied at $z \gtrsim 2$ due to the limited number of the background galaxies and their lower image quality. The second is the abundance matching. The abundance matching technique connects galaxies to their host dark matter haloes by matching the cumulative stellar mass function (or luminosity function) and the cumulative halo mass function. This technique is feasible even with a limited number of galaxies, because only one-point statistics is needed as an observable. However, one should make assumptions on a satellite galaxy-subhalo relation and a scatter in the central galaxy-host halo relation, which is related to the duty cycle (c.f., Reddick et al. 2013). The third is the clustering. In the clustering analysis, the clustering strength of galaxies is evaluated with the correlation function, and is compared to predictions of the $\Lambda$CDM structure formation models. Although the correlation strength is affected by the presence of satellite galaxies, the halo occupation distribution (HOD) model incorporates the effect of the satellite galaxies with the 1- and 2-halo terms. Thus the $M_h$ estimate is robust unless the applied structure formation models are wrong.

Some studies investigate the galaxy-dark matter halo connection at low redshift by combining these three methods. Leauthaud et al. (2012) study stellar-to-halo mass ratios (SHMRs) at $0.2 < z < 1.0$ using the COSMOS survey data, and find a redshift evolution of the SHMR. Since the SHMR com-prises the integrated efficiency of the past stellar mass assembly (i.e., star formation and mergers), the evolution of the SHMR would indicate an evolution of the galaxy formation processes with redshift. Coupon et al. (2015) investigate the SHMR of $M_\mathrm{h} \gtrsim 10^{12} \, M_\odot$ halos at $z \sim 0.8$ in the CFHTLenS/VIPERS field, whose large area reduces the cosmic variance. High-mass end of the galaxy-dark matter connection is also well explored by recent studies (e.g., Saito et al. 2016; Tinker et al. 2016). Some studies present SHMRs up to $z \sim 2.5$ by the clustering analysis (i.e., McCracken et al. 2015; Ishikawa et al. 2016a; Hatfield et al. 2016; Kashino et al. 2017).

Investigating the galaxy-dark matter halo connection at high redshift is also important, because a significant evolution of the galaxy formation processes is inferred from the downturn of the cosmic star formation rate density (SFRD) at $z \gtrsim 2$ (Madau & Dickinson 2014). Recently we identify a redshift evolution of the SHMR at $M_\mathrm{h} \sim 10^{11} \, M_\odot$ both from $z \sim 0$ to 4 and from $z \sim 4$ to 7 via our HOD modeling with Hubble Space Telescope and early Hyper-Suprime Cam Subaru strategic program (HSC-SSP) survey data (Harikane et al. 2016). However, the galaxy-halo connection is investigated only at the low mass halo of $4 \times 10^{10} \, M_\odot < M_\mathrm{h} < 3 \times 10^{11} \, M_\odot$ except at $z \sim 5$ with the HSC data (c.f. Hatfield et al. 2017). In addition, the satellite galaxy fraction is not constrained due to the small sample size of Harikane et al. (2016), although it is sensitive probe of star formation in subhalos (c.f., Ishikawa et al. 2016b). Abundance matching studies have probed the galaxy-dark matter halo connection in a wide halo mass range (e.g., $10^{11} \, M_\odot < M_\mathrm{h} < 10^{13} \, M_\odot$; Moster et al. 2013; Behroozi et al. 2013b) at $z \sim 0 - 8$, albeit with the uncertainties of the satellite galaxy and the scatter of the relation.

In this study, we investigate the galaxy-dark matter halo connection at $z \sim 4 - 6$ based on wide and deep optical HSC (Miyazaki et al. 2012; see also Miyazaki et al. 2017; Komiyama et al. 2017; Furusawa et al. 2017; Kawanomoto et al. 2017) images recently obtained by the HSC-SSP survey (Aihara et al. 2017b). Combined with the Hubble data from our previous study (Harikane et al. 2016) at $z \sim 4 - 7$, we can probe the connection over two orders of magnitude in the dark matter halo mass, $4 \times 10^{10} \, M_\odot < M_\mathrm{h} < 4 \times 10^{12} \, M_\odot$. A large number of galaxies found in the HSC data also allow us to constrain the redshift evolution of the galaxy-dark matter halo connection and the satellite fraction. This paper is one in a series of papers from twin programs devoted to scientific results on high redshift galaxies based on the HSC-SSP survey data. One program is our luminous Lyman break galaxy (LBG) studies, named Great Optically Luminous Dropout Research Using Subaru HSC (GOLDRUSH). It provides robust clustering measurements of luminous LBGs at $z \sim 4 - 6$, which are presented in this paper, the UV luminosity functions at $z \sim 4 - 7$ (Ono et al. 2017), and the correlation function of $z \sim 4$ galaxy proto-cluster candidates (Toshikawa et al. 2017). The other program is high redshift Ly$\alpha$ emitter studies using HSC narrowband filters, Systematic Identification of LAEs for Visible Exploration and Reionization Research Using Subaru HSC (SILVERRUSH; Ouchi et al. 2017; Shibuya et al. 2017a, 2017b; Konno et al. 2017, R. Higuchi et al. in preparation).

This paper is organized as follows. We show the observational data sets and describe sample selections in Section 2. The clustering analysis and results are presented in Section 3 and 4, respectively. We discuss our results in Section 5, and summarize our findings in Section 6. Throughout this paper we use the recent Planck cosmological parameter sets of the TT, TE, EE+lowP+lensing+ext result (Planck Collaboration et al. 2015): $\Omega_\mathrm{m} = 0.3089$, $\Omega_\Lambda = 0.6911$, $\Omega_\mathrm{b} = 0.049$, $h = 0.6774$, and $\sigma_8 = 0.8159$. We define $r_{200}$ that is the radius in which the mean enclosed density is 200 times higher than the mean cosmic density. To define the halo mass, we use $M_{200}$ that is the dark matter mass enclosed in $r_{200}$. Note that we used the total mass including both dark matter and baryons in our previous paper (Harikane et al. 2016). We assume a Chabrier (2003) initial mass function (IMF). All magnitudes are in the AB system (Oke & Gunn 1983).

# 2 Observational Data Sets and Sample Selection

## 2.1 HSC Data and Reduction

We use the internal data release product taken in the HSC-SSP survey (Aihara et al. 2017b) from 2014 March to 2016 April. Our data are larger than the first public data release in 2017 February, which includes the data taken in the first 1.7 years (2014 March to 2015 November) of the survey (Aihara et al. 2017a). Table 1 summarizes the HSC data used in this work. The HSC-SSP survey has three layers, the UltraDeep, Deep, and Wide, with different combinations of area and depth. Total effective survey areas of the Ultra-Deep, Deep, and Wide layers are $\sim 2$, $\sim 18$, and $\sim 83 \, \mathrm{deg}^2$, respectively. Here we define the effective survey area as area where the number of visits in $g$, $r$, $i$, $z$, and $y$-bands are equal or larger than threshold values after masking interpolated, saturated, or bad pixels, cosmic rays, and bright source halos (Coupon et al. 2017). The applied flags are summarized in Table 2. The threshold values are $(13, 13, 27, 41, 38)$, $(17, 16, 27, 47, 62)$, and $(3, 3, 5, 5, 5)$ for $(g, r, i, z, y)$ for UD-SXDS, UD-COSMOS, and other fields, respectively. In addition to these flags, we mask some regions in the Wide layers with poor PSF modeling due to too good seeing (see Aihara et al. 2017a). The regions used in this study are presented in Figure 1-9. Typical $r$-band $5\sigma$ limiting magnitudes measured in a $1\farcs5$-diameter circular aperture are 26.8, 26.2, and 25.9 mag in the UltraDeep, Deep, and Wide layers, respectively, which are estimated later.

**Table 1.** HSC SSP Data Used in This Study

| Field | R.A. (J2000) | decl (J2000) | Area† (deg²) | $g$ (ABmag) | $r$ (ABmag) | $i$ (ABmag) | $z$ (ABmag) | $y$ (ABmag) |
|-------|--------------|--------------|--------------|-------------|-------------|-------------|-------------|-------------|
| (1) | (2) | (3) | (4) | (5) | (6) | (7) | (8) | (9) |
| | | | UltraDeep | | | | | |
| UD-SXDS | 02:18:00.00 | -05:00:00.00 | 1.11 | 27.15 | 26.68 | 26.53 | 25.96 | 25.15 |
| UD-COSMOS | 10:00:28.60 | 02:12:21.00 | 1.29 | 27.13 | 26.84 | 26.46 | 26.10 | 25.28 |
| | | | Deep | | | | | |
| D-XMM-LSS | 02:16:51.57 | -03:43:08.43 | 2.37 | 26.73 | 26.30 | 25.88 | 25.42 | 24.40 |
| D-COSMOS | 10:00:59.50 | 02:13:53.06 | 6.53 | 26.56 | 26.56 | 26.04 | 25.58 | 24.76 |
| D-ELAIS-N1 | 16:10:00.00 | 54:17:51.07 | 3.33 | 26.77 | 26.13 | 25.87 | 25.16 | 24.25 |
| D-DEEP2-3 | 23:30:22.22 | -00:44:37.69 | 5.53 | 26.69 | 26.25 | 25.96 | 25.29 | 24.56 |
| | | | Wide | | | | | |
| W-XMM | 02:16:51.57 | -03:43:08.43 | 28.45 | 26.43 | 25.93 | 25.71 | 25.00 | 24.25 |
| W-GAMA09H | 09:05:11.11 | 00:44:37.69 | 12.36 | 26.35 | 25.88 | 25.65 | 25.07 | 24.45 |
| W-WIDE12H | 11:57:02.22 | 00:44:37.69 | 15.20 | 26.38 | 25.95 | 25.82 | 25.15 | 24.23 |
| W-GAMA15H | 14:31:06.67 | -00:44:37.69 | 16.57 | 26.39 | 25.96 | 25.81 | 25.11 | 24.31 |
| W-HECTOMAP | 16:08:08.14 | 43:53:03.47 | 4.81 | 26.47 | 26.04 | 25.82 | 25.09 | 24.07 |
| W-VVDS | 22:37:02.22 | 00:44:37.69 | 5.14 | 26.31 | 25.87 | 25.74 | 24.98 | 24.23 |
| Total | — | — | 102.69 | | | | | |

Columns: (1) Field name. (2) Right ascension. (3) Declination. (4) Effective area in deg², (5)-(9) $5\sigma$ limiting magnitude measured with $1\rlap{.}''5$ diameter circular apertures in $g$, $r$, $i$, $z$, and $y$.
† The effective area may be smaller than the full data sets, because we mask some regions in the Wide layers with poor PSF modeling due to too good seeing (Aihara et al. 2017a). See Figure 1-9 for the regions used in this study

**Table 2.** The Selection Criteria for Our Catalog Construction

| Parameter | Value | Band | Comment |
|-----------|-------|------|---------|
| detect_is_primary | True | — | Object is a primary one with no deblended children |
| flags_pixel_edge | False | $grizy$ | Locate within images |
| flags_pixel_interpolated_center | False | $grizy$ | None of the central $3 \times 3$ pixels of an object is interpolated |
| flags_pixel_saturated_center | False | $grizy$ | None of the central $3 \times 3$ pixels of an object is saturated |
| flags_pixel_cr_center | False | $grizy$ | None of the central $3 \times 3$ pixels of an object is masked as cosmic ray |
| flags_pixel_bad | False | $grizy$ | None of the pixels in the footprint of an object is labelled as bad |
| flags_pixel_bright_object_any | False | $grizy$ | None of the pixels in the footprint of an object is close to bright sources |
| centroid_sdss_flags | False | $ri$ for $g$-drop | Object centroid measurement has no problem |
| | False | $iz$ for $r$-drop | — |
| | False | $zy$ for $i$-drop | — |
| cmodel_flux_flags | False | $gri$ for $g$-drop | Cmodel flux measurement has no problem |
| | False | $riz$ for $r$-drop | — |
| | False | $izy$ for $i$-drop | — |
| merge_peak | True | $ri$ for $g$-drop | Detected in $r$ and $i$. |
| | False/True | $g/iz$ for $r$-drop | Undetected in $g$ and detected in $r$ and $i$. |
| | False/True | $gr/zy$ for $i$-drop | Undetected in $g$ and $r$, and detected in $z$ and $y$. |
| blendedness_abs_flux | < 0.2 | $ri$ for $g$-drop | The target photometry is not significantly affected by neighbors. |
| | < 0.2 | $iz$ for $r$-drop | — |
| | < 0.2 | $zy$ for $i$-drop | — |

The HSC data are reduced by the HSC SSP collaboration with hscPipe (Bosch et al. 2017) that is the HSC data reduction pipeline based on the Large Synoptic Survey Telescope (LSST) pipeline (Ivezic et al. 2008; Axelrod et al. 2010). hscPipe performs all the standard procedures including bias subtraction, flat fielding with dome flats, stacking, astrometric and photometric calibrations, source detections and measurements, and construction of a multiband photometric catalog. The astrometric and photometric calibration are based on the data of Panoramic

Survey Telescope and Rapid Response System (Pan-STARRS) 1 imaging survey (Magnier et al. 2013; Schlafly et al. 2012; Tonry et al. 2012). PSFs are calculated in hscPipe (Jee & Tyson 2011), and typical FWHMs of the PSFs are $0\rlap{.}''6 - 0\rlap{.}''9$.

We use forced photometry, which allows us to measure fluxes in multiple bands with a consistent aperture defined in a reference band. The reference band is $i$ by default and is switched to $z$ ($y$) for sources with no detection in the $i$ ($z$) and bluer bands. We measure total fluxes and colors of sources with the cModel magnitude, $m_{cModel}$, which is measured by fitting PSF-convolved galaxy models to the source profile (Abazajian et al. 2004). Limiting magnitudes and source detections are evaluated with aperture magnitudes, $m_{aper}$. All the magnitudes are corrected for Galactic extinction (Schlegel et al. 1998).

We measure the $5\sigma$ limiting magnitudes which are defined as the $5\sigma$ levels of sky noise in a $1''5$ diameter aperture. This definition is not the same as that in the data release paper of Aihara et al. (2017a), who use the pipeline outputs of flux errors. Since the error outputs of hscPipe would be underestimated as discussed in Aihara et al. (2017a), we use the classical definition of the limiting magnitude. The sky noise is calculated from the fluxes in sky apertures which are randomly placed on the image in the reduction process. The limiting magnitudes measured in $g$, $r$, $i$, $z$, and $y$-bands are presented in Table 1.

## 2.2 Sample Selection

In this work, we investigate the clustering of $z \sim 4 - 6$ galaxies. We construct high redshift galaxy samples with the LBG selection method (e.g., Steidel et al. 1996; Giavalisco 2002). Many photometric and spectroscopic observations have revealed that LBGs are star-forming galaxies (e.g., Shapley et al. 2001; Erb et al. 2006) with emission and/or absorption lines in rest-frame UV (e.g., Shapley et al. 2003; Le Fèvre et al. 2015), galactic outflows (e.g., Erb et al. 2012), and various dust contents (e.g., Reddy et al. 2016). Here we describe a summary of our LBG selection method, which is the same as Ono et al. (2017). We select LBGs at $z \sim 4$, 5, and 6 from our multi-band photometric catalog by applying following color selection criteria:

$z \sim 4$

$$g - r > 1.0, \tag{1}$$
$$r - i < 1.0, \tag{2}$$
$$g - r > 1.5(r - i) + 0.8, \tag{3}$$

$z \sim 5$

$$r - i > 1.2, \tag{4}$$
$$i - z < 0.7, \tag{5}$$
$$r - i > 1.5(i - z) + 1.0, \tag{6}$$

$z \sim 6$

$$i - z > 1.5, \tag{7}$$
$$z - y < 0.5, \tag{8}$$
$$i - z > 2.0(z - y) + 1.1. \tag{9}$$

We select galaxies that have Lyman breaks according to the criteria of Equations (1), (4), and (7), and exclude intrinsically-red galaxies by the additional constraints of Equations (2), (3), (5), (6), (8), and (9). The selection criteria at $z \sim 4$ and 5 are similar to those of the CFHT study (Hildebrandt et al. 2009), which

**Table 3.** Number of LBGs in Our $z \sim 4$, 5, and 6 Samples

| Field | $z \sim 4$ | $z \sim 5$ | $z \sim 6$ |
|---|---|---|---|
| (1) | (2) | (3) | (4) |
| *UltraDeep* | | | |
| UD-SXDS | 9916 | 1209 | 36 |
| UD-COSMOS | 10644 | 1990 | 50 |
| *Deep* | | | |
| D-XMM-LSS | 6730 | 711 | 6 |
| D-COSMOS | 45767 | 6282 | 64 |
| D-ELAIS-N1 | 19631 | 612 | 15 |
| D-DEEP2-3 | 35963 | 1498 | 47 |
| *Wide* | | | |
| W-XMM | 113582 | 6371 | 81 |
| W-GAMA09H | 44670 | 5989 | 98 |
| W-WIDE12H | 94544 | 5243 | 36 |
| W-GAMA15H | 104224 | 6457 | 73 |
| W-HECTOMAP | 30663 | 1082 | 11 |
| W-VVDS | 23677 | 1500 | 20 |
| $N_{total}(z)$ | 540011 | 38944 | 537 |
| $N_{total}$ | | 579492 | |

Columns: (1) Field. (2)-(4) Number of LBGs at each redshift.

uses the photometric system almost identical to the one of our HSC data.

In addition to the color selection criteria, we adopt the following three criteria. First, to identify secure sources, we apply detection limit of $> 5\sigma$ levels in the $i$, $z$, and $z$-bands and require secure measurements of centroid positions in the $ri$ and $iz$, $zy$-bands at $z \sim 4$, 5, and 6, respectively (Table 2). In addition, we require a $4\sigma$ detection in the $y$-band for the $z \sim 6$ selection, since the $y$-band image is slightly shallow. Second, for reducing foreground interlopers, we exclude sources with continuum detection at $> 2\sigma$ levels in $g$-band data for the $z \sim 5$ selection and $g$- or $r$-bands for the $z \sim 6$ selection. Third, to remove severely blended sources, we apply a blendedness parameter threshold of $b < 0.2$ in $ri$, $iz$, and $zy$-bands at $z \sim 4$, 5, and 6, respectively (Table 2). This blendedness parameter is defined as $b = 1 - f_T / f_{T+N}$, where $f_T$ and $f_{T+N}$ are fluxes of the target and the sum of target and neighbors, respectively. After adopting these criteria, the contamination fraction is very small, i.e., less than 10% for the $z \sim 4$ and 5 LBG samples in the Deep layers based on simulations (Ono et al. 2017).

We construct a total sample of 540,011, 38,944, and 537 LBGs at $z \sim 4$, 5, and 6, respectively[1]. Our sample is selected from the $102.69 \ deg^2$ wide area data corresponding to a $1.39 \ Gpc^3$ survey volume, and is the largest sample of the high-redshift ($z \gtrsim 4$) galaxy population to date. Table 3 summarizes the number of LBGs in each field, and Figure 1-9 show sky distributions of the LBGs. The number of the $z \sim 6$ LBGs

---

[1] Machine-readable catalogs of the LBGs will be provided on our project webpage at http://cos.icrr.u-tokyo.ac.jp/rush.html.

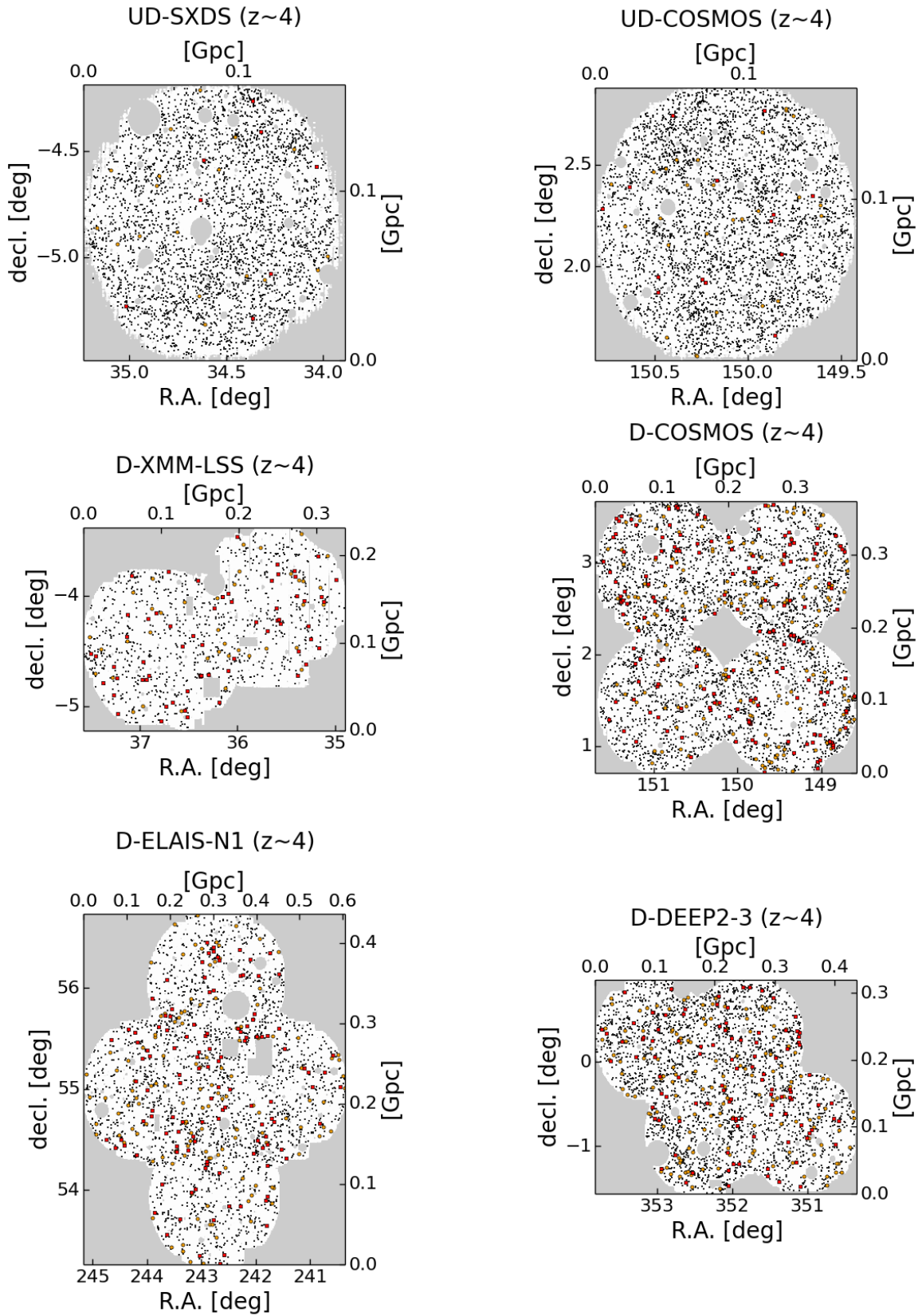

**Fig. 1.** Sky distribution of $z \sim 4$ LBGs in the UltraDeep and Deep layers. The red squares, orange circles, and black dots represent positions of LBGs whose magnitudes are $i < 23.0\,\mathrm{mag}$, $23.0 - 23.5\,\mathrm{mag}$, and $23.5 - 26.0(25.0)\,\mathrm{mag}$ in the UltraDeep (Deep) layers, respectively. The scale on the map is marked in degrees and the projected distance (comoving Gpc).

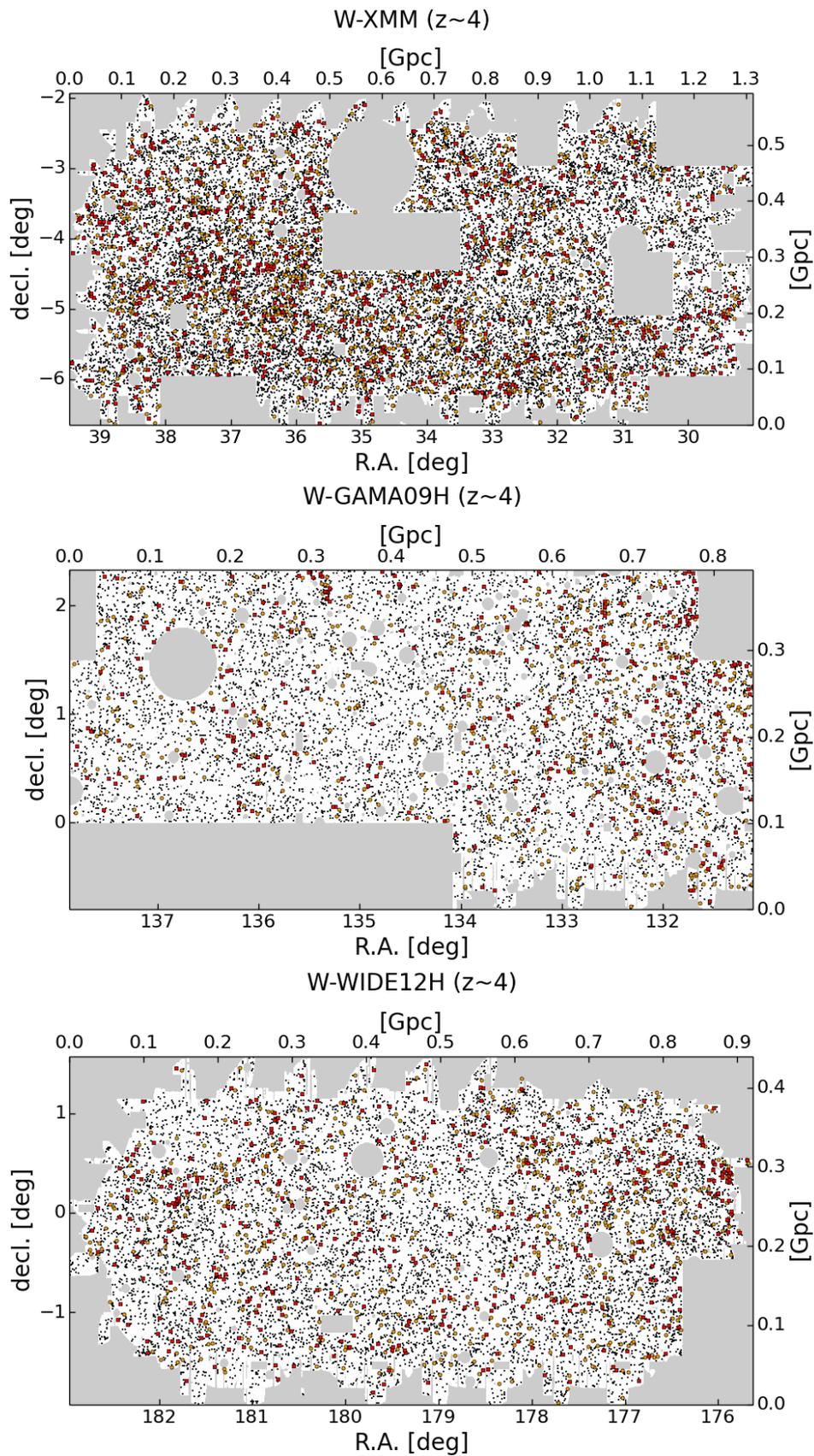

**Fig. 2.** Same as Figure 1 but in the Wide layers. The red squares, orange circles, and black dots represent positions of LBGs whose magnitudes are $i < 23.0 \, \mathrm{mag}$, $23.0 - 23.5 \, \mathrm{mag}$, and $23.5 - 25.0 \, \mathrm{mag}$ in the Wide layers, respectively.

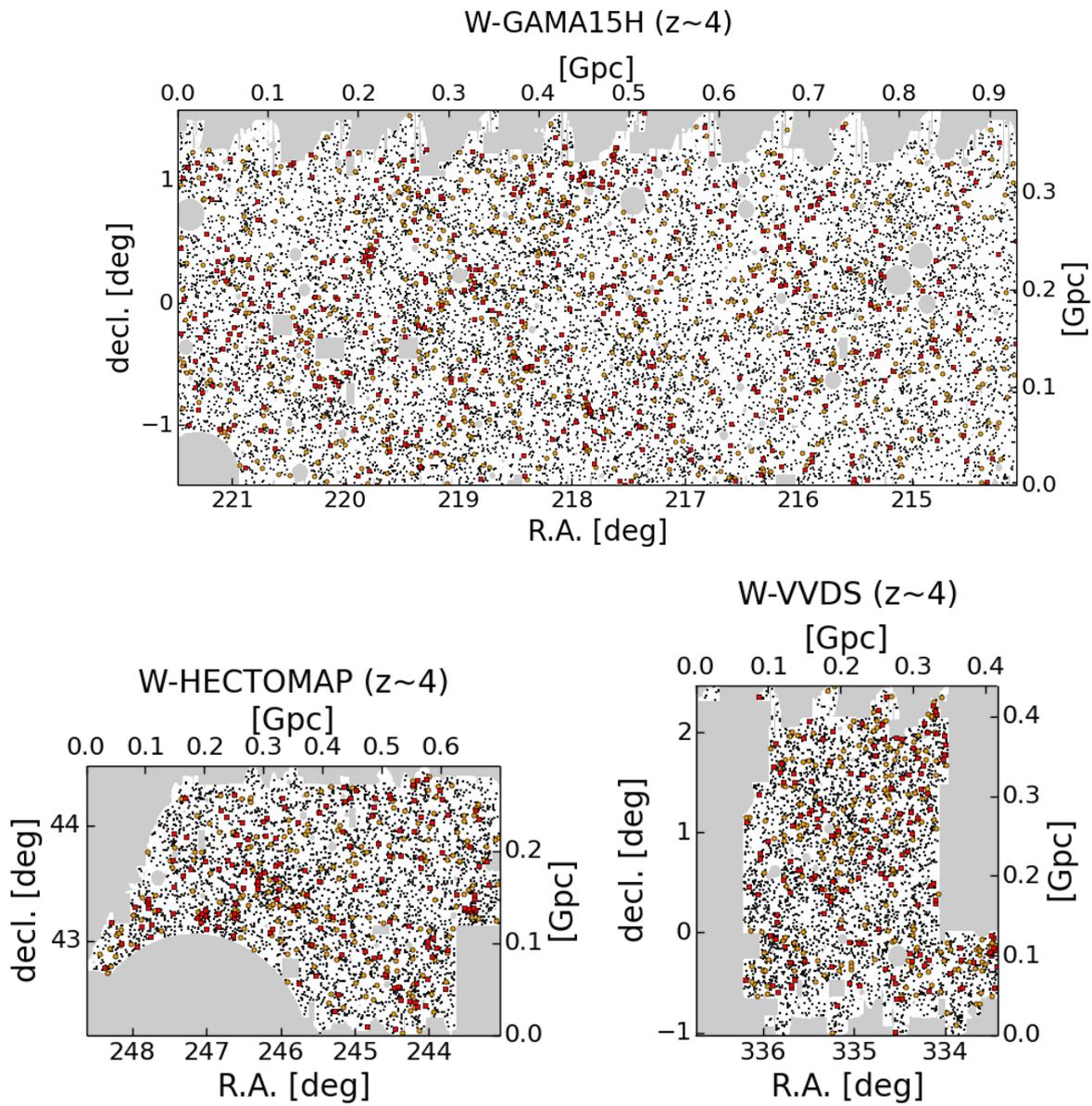

**Fig. 3.** Cotinued.

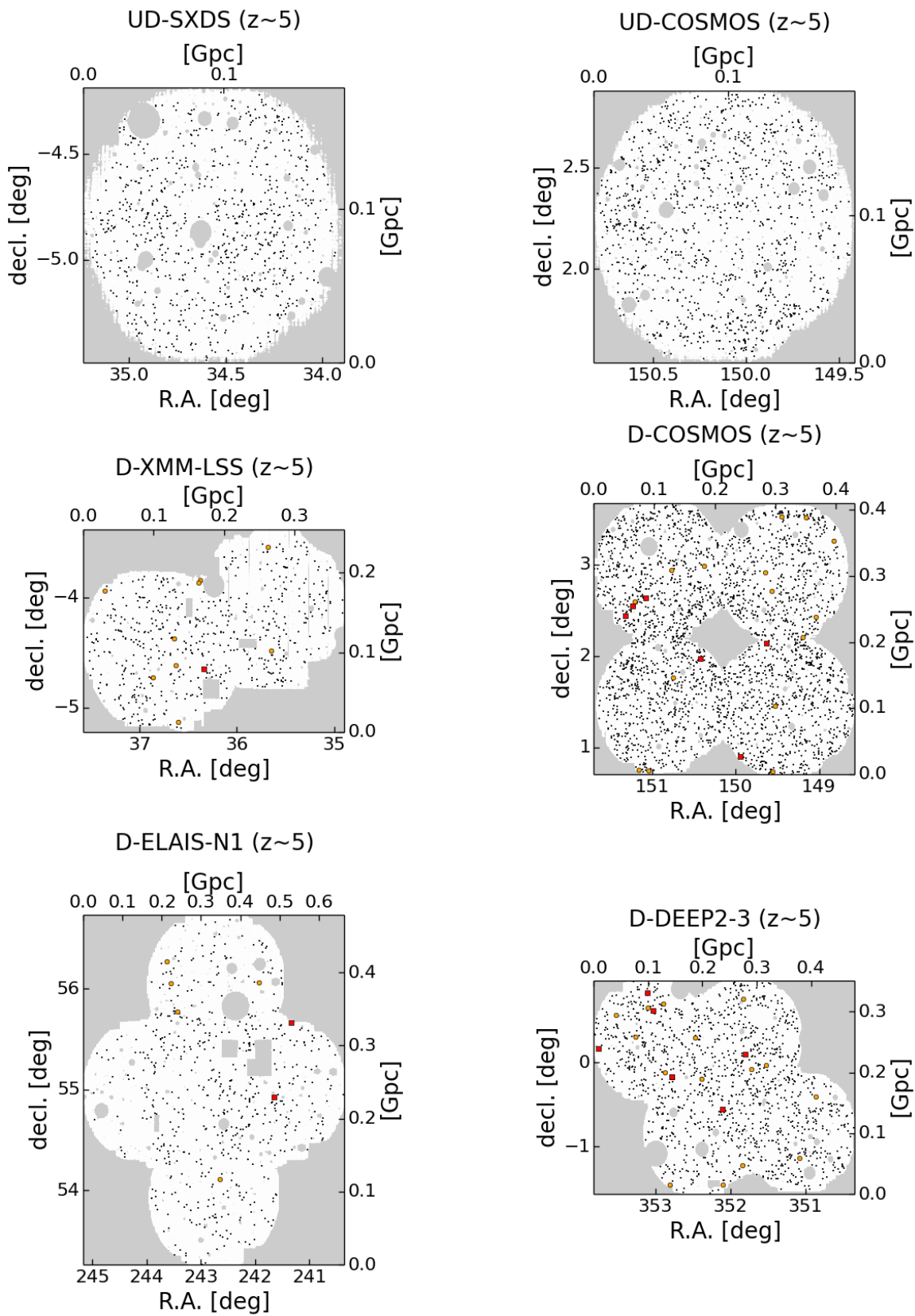

**Fig. 4.** Sky distribution of $z \sim 5$ LBGs in the UltraDeep and Deep layers. The red squares, orange circles, and black dots represent positions of LBGs whose magnitudes are $z < 23.0$ mag, $23.0 - 23.5$ mag, and $23.5 - 26.0(25.0)$ mag in the UltraDeep (Deep) layers, respectively. The scale on the map is marked in degrees and the projected distance (comoving Gpc).

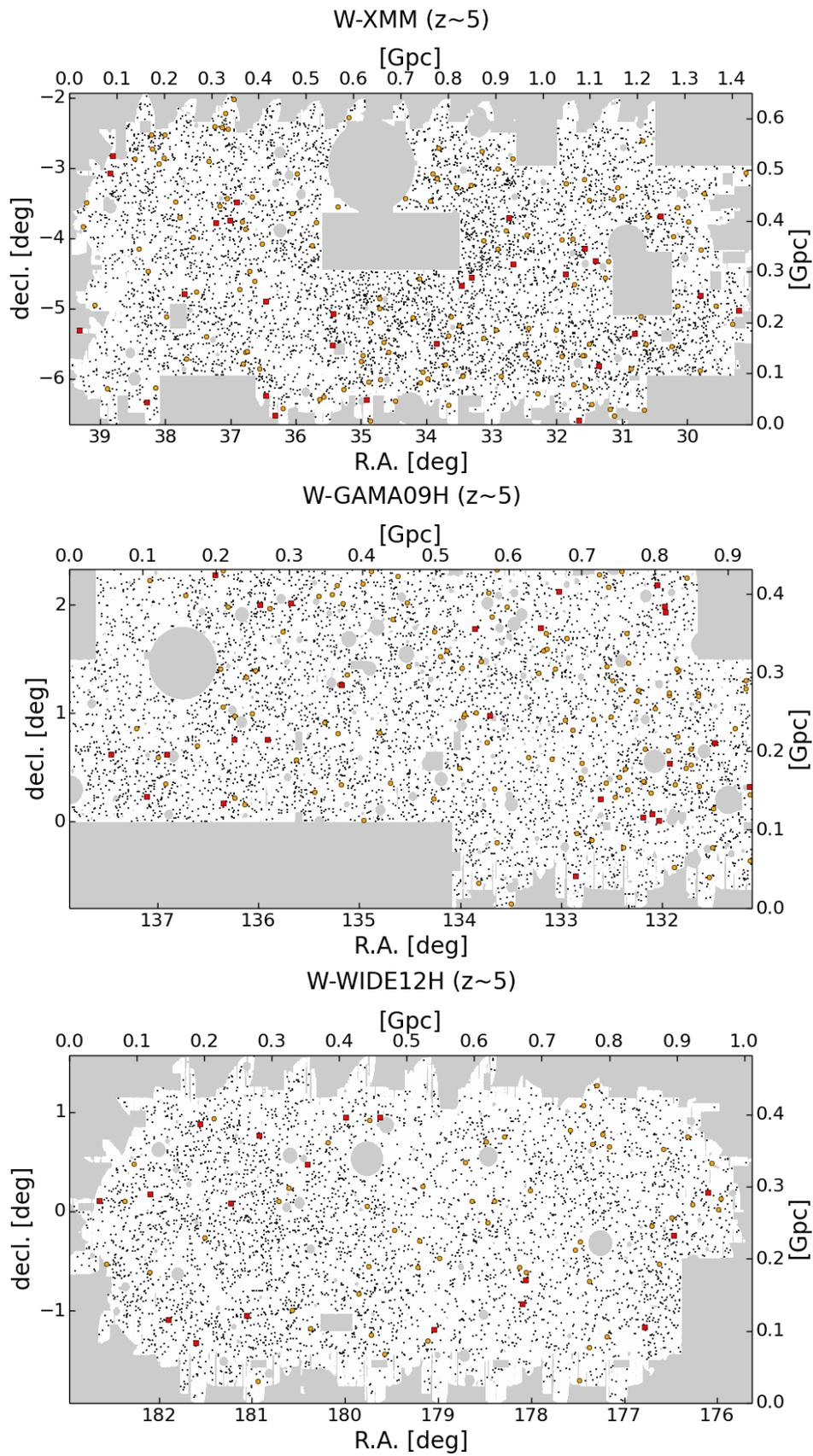

**Fig. 5.** Same as Figure 4 but in the Wide layers. The red squares, orange circles, and black dots represent positions of LBGs whose magnitudes are $z < 23.0\,\mathrm{mag}$, $23.0 - 23.5\,\mathrm{mag}$, and $23.5 - 25.0\,\mathrm{mag}$ in the Wide layers, respectively.

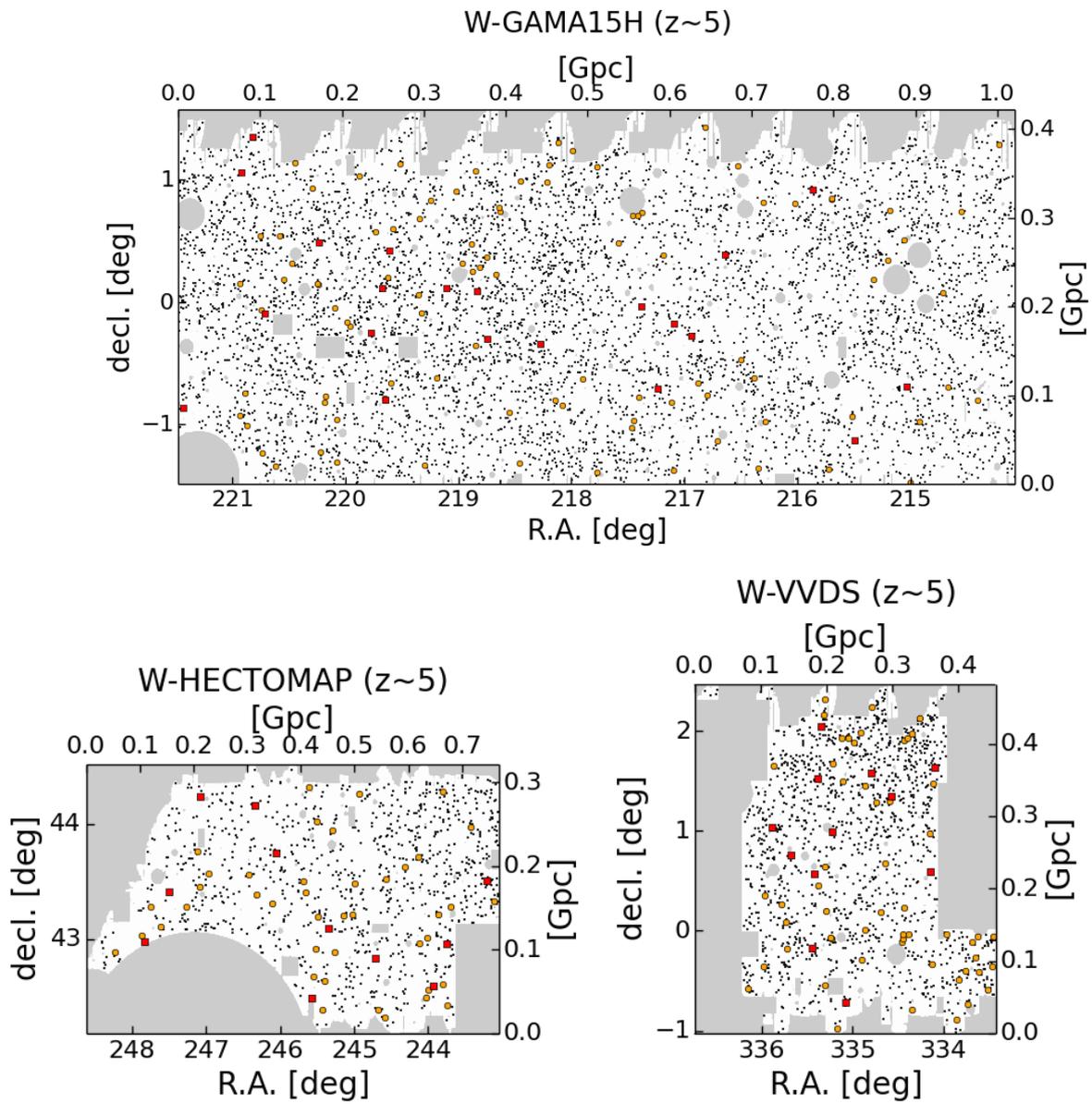

**Fig. 6.** Cotinued.

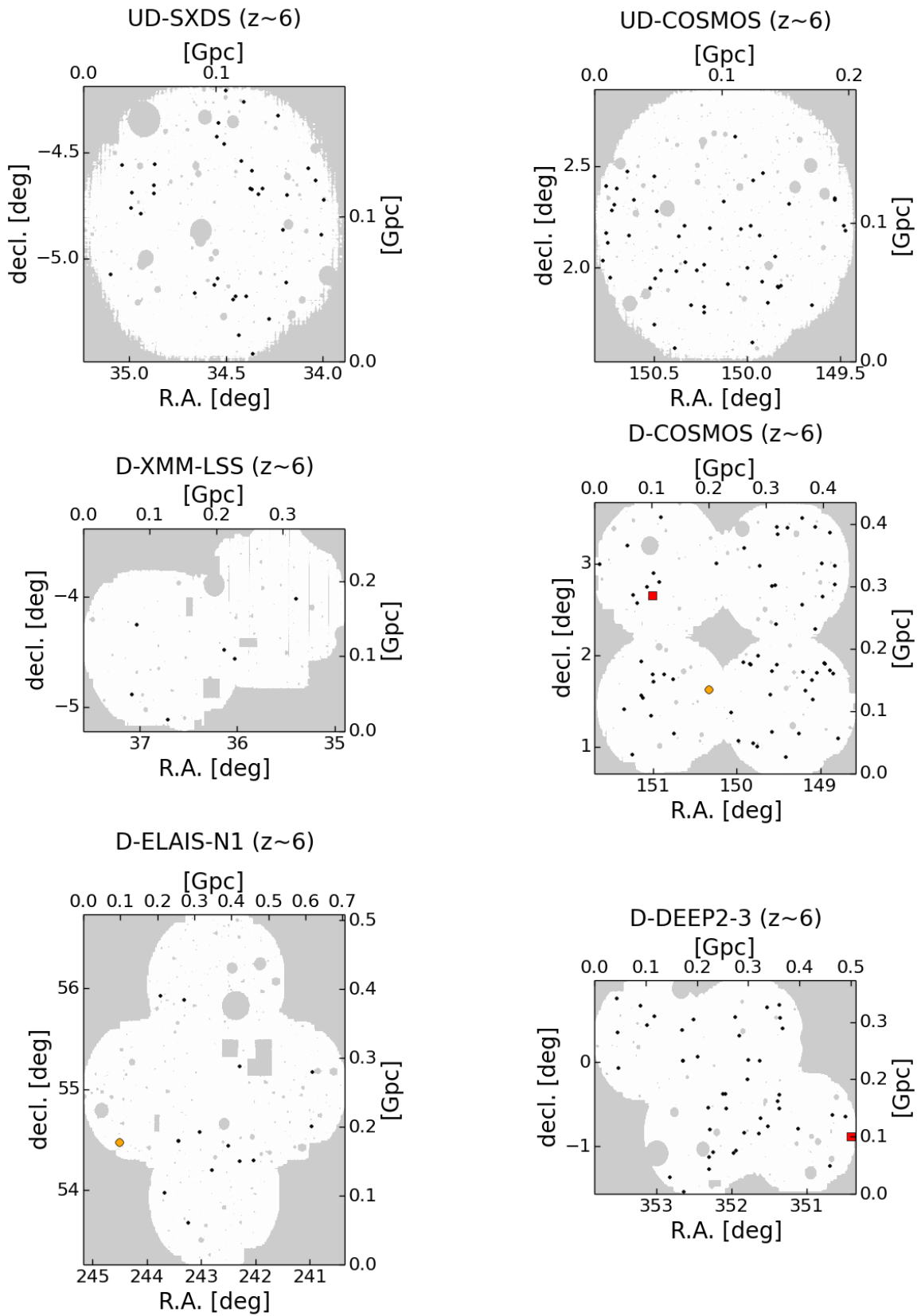

**Fig. 7.** Sky distribution of $z \sim 6$ LBGs in the UltraDeep and Deep layers. The red squares, orange circles, and black dots represent positions of LBGs whose magnitudes are $y < 23.0 \, \mathrm{mag}$, $23.0 - 23.5 \, \mathrm{mag}$, and $23.5 - 26.0(25.0) \, \mathrm{mag}$ in the UltraDeep (Deep) layers, respectively. The scale on the map is marked in degrees and the projected distance (comoving Gpc).

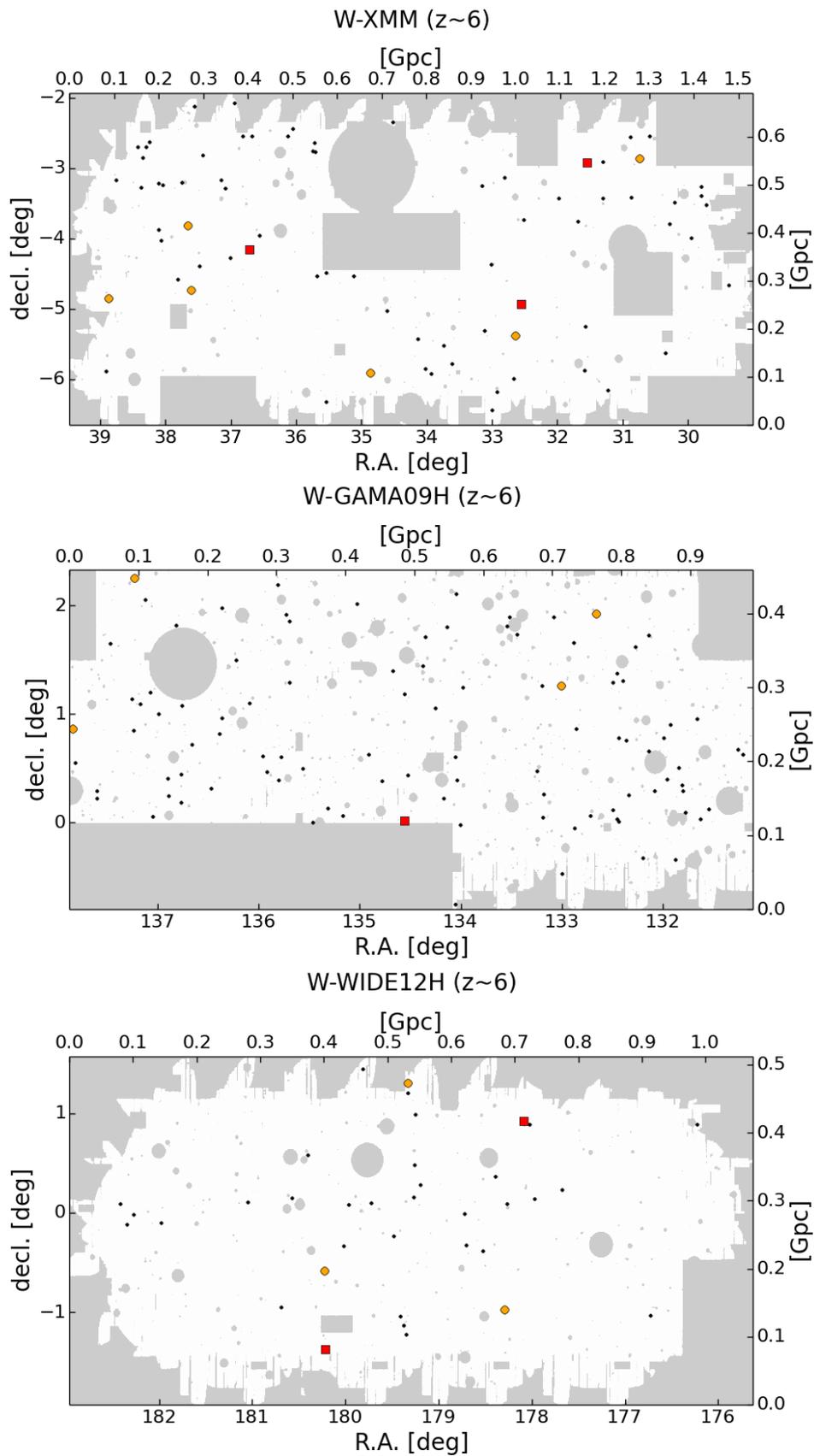

**Fig. 8.** Same as Figure 7 but in the Wide layers. The red squares, orange circles, and black dots represent positions of LBGs whose magnitudes are $y < 23.0$ mag, $23.0 - 23.5$ mag, and $23.5 - 25.0$ mag in the Wide layers, respectively. The scale on the map is marked in degrees and the projected distance (comoving Gpc).

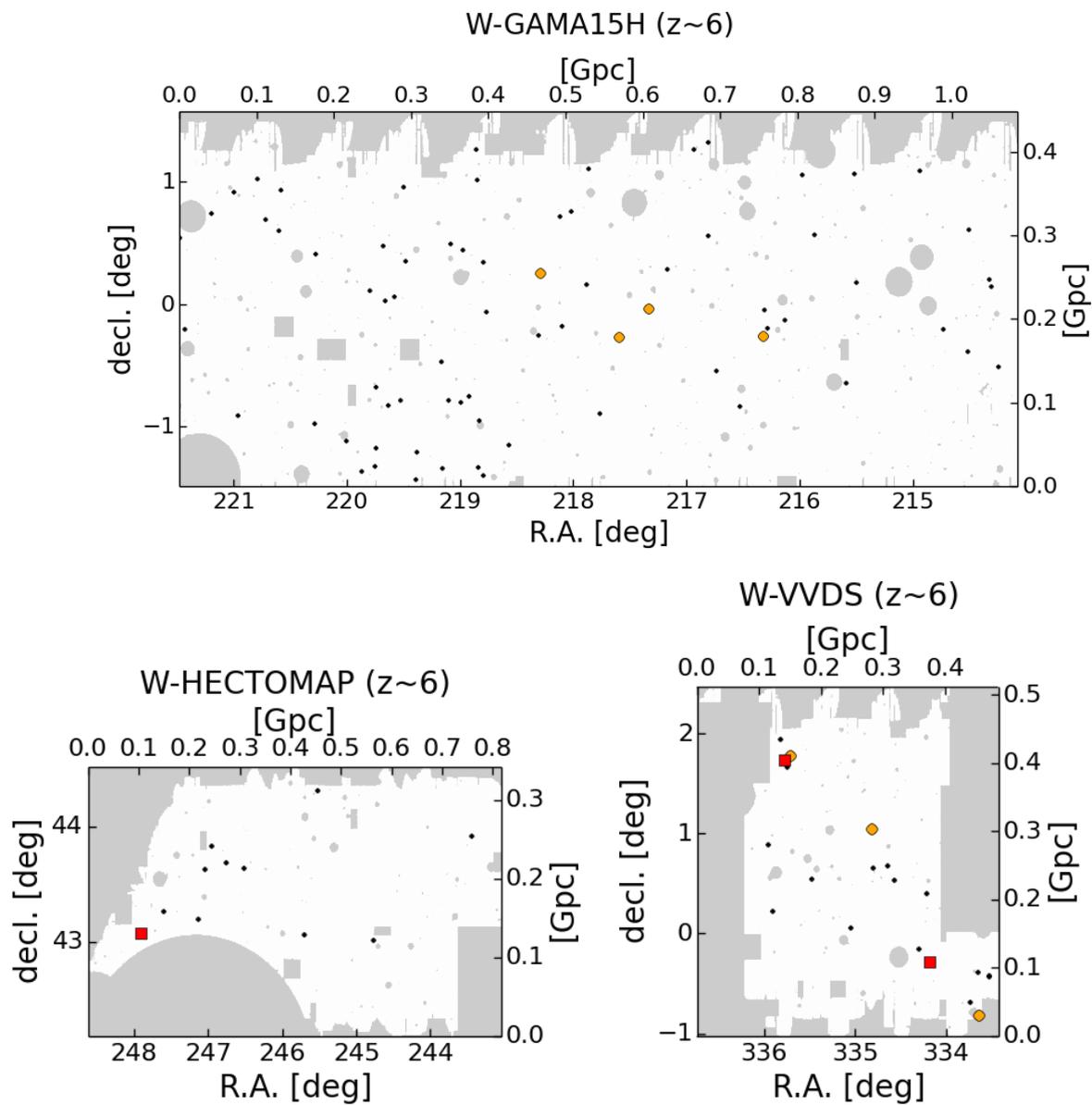

**Fig. 9.** Cotinued.

is slightly small due to the shallow depth in the $y$-band images. The selection completeness and redshift distributions of our LBGs are estimated by Monte Carlo simulations (Ono et al. 2017). The mean redshifts of the $z \sim 4$, 5, and 6 LBGs are $z_c = 3.8$, 4.9, and 5.9, respectively. We define the UV magnitude, $m_{UV}$, as the cModel magnitude in the band whose central wavelength is the nearest to the rest-frame wavelength of 1500 Å. We do not use objects brighter than $m_{UV}^{cut} = 23.5$ mag at each redshift in our analysis, since the quasar fraction is moderately high based on spectroscopic observations (Ono et al. 2017; Matsuoka et al. 2016, 2017). Changing this cut to brighter magnitude does not affect our conclusions. For reference, we also show results with the magnitude cut of $m_{UV}^{cut} = 20.0$ mag instead of 23.5 mag (see Table 4). The obtained results of $m_{UV}^{cut} = 20.0$ mag agree with those of $m_{UV}^{cut} = 23.5$ mag within the $\sim 1\sigma$ errors.

## 3 Clustering Analysis

### 3.1 Angular Correlation Function (ACF)

To test the dependence of the clustering strength on the luminosity, we divide our LBG sample into subsamples by UV magnitude thresholds ($m_{UV}^{th}$). The number of LBGs in the subsamples and their magnitude thresholds are summarized in Table 4. The number of LBGs used in the subsamples is smaller than that of our entire sample, because we use LBGs of $m_{UV} \leq 25.5$ mag ($m_{UV} \leq 24.0$ mag) in the UltraDeep (Deep and Wide) layers where the selection completeness is larger than $\sim 80\%$ at $z \sim 4$, and 5.

We calculate observed ACFs, $\omega_{obs}(\theta)$, with the subsamples, using an estimator proposed by Landy & Szalay (1993),

$$\omega_{obs}(\theta) = \frac{DD(\theta) - 2DR(\theta) + RR(\theta)}{RR(\theta)}, \quad (10)$$

where $DD(\theta)$, $DR(\theta)$, and $RR(\theta)$ are the numbers of galaxy-galaxy, galaxy-random, and random-random pairs normalized by the total numbers of pairs. We use the random catalog whose surface number density is $100 \, \mathrm{arcmin}^{-2}$ with the same geometrical shape as the observational data including the mask positions (Coupon et al. in prep.). We calculate ACFs in individual fields, and obtain the best-estimate ACF which is the weighted mean by the effective area in each subsample.

Due to the finite size of our survey fields, the observed ACF is underestimated by a constant value known as the integral constraint, $IC$ (Groth & Peebles 1977). Including a correction for the number of objects in the sample, $N$ (Peebles 1980), the true ACF is given by

$$\omega(\theta) = \omega_{obs}(\theta) + IC + \frac{1}{N}. \quad (11)$$

We estimate the integral constraint with

$$IC = \frac{\Sigma_i RR(\theta_i) \omega_{model}(\theta_i)}{\Sigma_i RR(\theta_i)}, \quad (12)$$

where $\omega_{model}(\theta)$ is the best-fit model ACF, and $i$ refers the angular bin. $IC$ and $\omega_{model}(\theta)$ are determined simultaneously in the HOD model fitting in Section 3.3.

We estimate statistical errors of the ACFs using the Jackknife estimator. We divide each subsample into Jackknife samples of about $1000^2 \, \mathrm{arcsec}^2$, whose size is larger than the largest angular scale in the ACFs. Removing one Jackknife sample at a time for each realization, we compute the covariance matrix as

$$C_{ij} = \frac{N-1}{N} \sum_{l=1}^{N} \left[\omega^l(\theta_i) - \bar{\omega}(\theta_i)\right] \left[\omega^l(\theta_j) - \bar{\omega}(\theta_j)\right]. \quad (13)$$

where $N$ is the total number of the Jackknife samples, and $\omega^l$ is the estimated ACF from the $l$th realization. $\bar{\omega}$ denotes the mean ACF. The total number of the Jackknife sample is $N = 2012$ (54) for the $z \sim 4 - 5$ $m_{UV}^{th} = 24.0$, 24.5 and $z \sim 6$ $m_{UV}^{th} = 25.0$ ($z \sim 4 - 5$ $m_{UV}^{th} = 25.0$, 25.5) subsamples. We apply a correction factor given by Hartlap et al. (2007) to an inverse covariance matrix in order to compensate for the bias introduced by the noise.

### 3.2 Test of Our ACF Estimates with a Simulation

We compare our ACFs based on the `hscPipe` photometry with literature derived with SExtractor (Bertin & Arnouts 1996) in Figure 10. Our ACFs are in excellent agreement with those of Hildebrandt et al. (2009) and Ouchi et al. (2005) at the scale of $\theta > 10''$. On the other hand, our ACFs are lower than those of literature at the small scale of $\theta < 10''$. This difference could be caused by low detection and/or selection completeness of the close ($\theta < 10''$) LBG pairs with `hscPipe`.

In order to investigate whether these lower ACFs are real signals or not, we calculate detection and selection completeness of galaxy pairs as functions of the pair separation by running a Monte Carlo simulation with an input mock catalog of LBGs. In the mock catalog, a size distribution and the Sersic index of LBGs follow results of Shibuya et al. (2015). We assume a uniform distribution of the intrinsic ellipticities in the range of $0.0 - 0.8$, since the observational results of $z \sim 3 - 5$ LBGs have roughly uniform distributions (Ravindranath et al. 2006). Position angles are randomly chosen. We use the stellar population synthesis model of GALAXEV (Bruzual & Charlot 2003) to produce galaxy spectra. We adopt the Salpeter (1955) IMF with lower and upper mass cutoffs of 0.1 $M_\odot$ and 100 $M_\odot$, a constant rate of star formation, age of 25 Myr, metallicity of $Z/Z_\odot = 0.2$, and Calzetti et al. (2000) dust extinction ranging from $E(B - V) = 0.0 - 0.4$. The IGM absorption is taken into account by using a prescription of Madau (1995).

We carry out the simulation with `SynPipe` (Huang et al. 2017). We insert 205,920 mock $z \sim 4$ LBGs into real HSC images of individual CCDs at a single exposure level. We then stack the single exposure images, create a source cata-log, and select LBGs in the same manner as in Section 2.2. Then we calculate the detection and selection completeness of LBG pairs as functions of the pair separation. Figure 11 presents the results for $z \sim 4$ LBG pairs whose magnitudes are $23.0 \; \mathrm{mag} < i < 25.0 \; \mathrm{mag}$. The detection completeness is about 80% and does not depend on the separation. The selection completeness drops at the very small separation of $\theta < 2''$, but is almost constant at $\theta > 2''$. In addition, the pair selection completeness at $\theta > 2''$ ($\sim 30\%$) is consistent with the squared completeness for the single LBG of $23.0 \; \mathrm{mag} < i < 25.0 \; \mathrm{mag}$ ($P^2_{\mathrm{single\,sel.}} = 0.55^2$; Ono et al. 2017). These results indicate that our ACF measurements with hscPipe are robust at $\theta \geq 2''$, and the lower ACFs at $2'' < \theta < 10''$ in Figure 10 cannot be explained by the detection or selection completeness. Thus in this paper, we conclude that the lower ACFs are real signals. Simulations with SExtractor (Bertin & Arnouts 1996) used in Hildebrandt et al. (2009) and Ouchi et al. (2005) are needed for the further discussion.

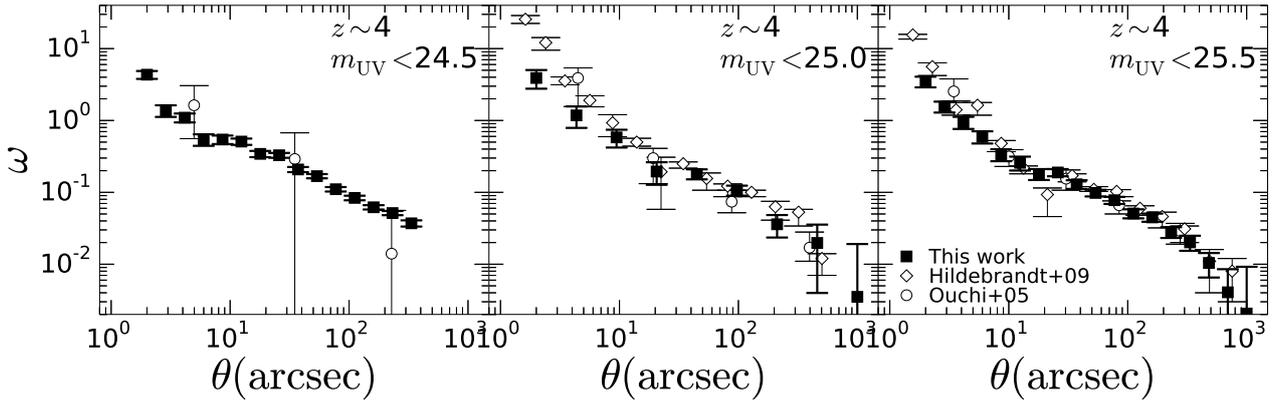

**Fig. 10.** Comparisons with the ACFs in the literature. The black squares represent the ACFs in this study. The open diamonds and circles denote ACFs in Hildebrandt et al. (2009) and Ouchi et al. (2005), respectively. Our ACFs agree well with those of Hildebrandt et al. (2009) and Ouchi et al. (2005) at the scale of $\theta > 10''$, while are smaller than those of previous studies at the small scale of $\theta < 10''$ (see Section 3.2).

### 3.3 HOD Model

We use an HOD model to derive estimates of the average dark matter halo properties for our selected galaxy samples. The HOD model is an analytic framework quantifying a probability distribution for the number of galaxies in the dark matter halos. The key assumption in the HOD model is that the probability depends only on the halo mass, $M_{\mathrm{h}}$. We can analytically calculate ACFs and number densities from the HOD model. Details of the calculations are presented in Harikane et al. (2016).

We fit our HOD model to the observed ACFs and number densities. In the fitting procedures, the best-fit parameters are determined by minimizing the $\chi^2$ value,

$$\chi^2 = \sum_{i,j} [\omega(\theta_i) - \omega_{\mathrm{model}}(\theta_i)] \, C^{-1}_{i,j} \, [\omega(\theta_j) - \omega_{\mathrm{model}}(\theta_j)]$$

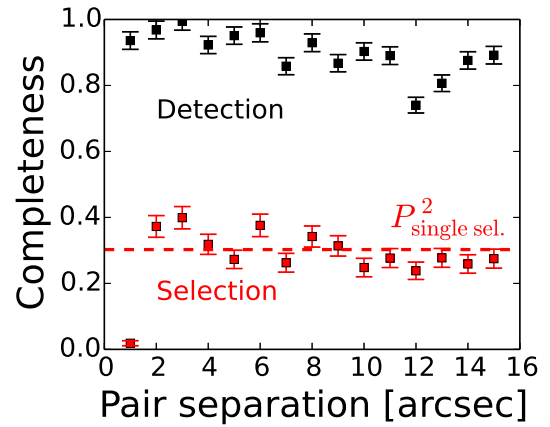

**Fig. 11.** Detection and selection completeness of galaxy pairs by the simulation. The detection and selection completeness are represented by black and red squares, respectively, as functions of the pair separation. The selection completeness at $\theta \geq 2''$ is almost constant, indicating that our ACF estimates are robust in this range. The red dashed line denotes the squared completeness for the single source of $23.0 \; \mathrm{mag} < i < 25.0 \; \mathrm{mag}$ ($P^2_{\mathrm{single\,sel.}} = 0.55^2$), which is consistent with the selection completeness of the pairs at $\theta \geq 2''$ ($\sim 30\%$).

$$+ \frac{[\log n^{\mathrm{obs}}_{\mathrm{g}} - \log n^{\mathrm{model}}_{\mathrm{g}}]^2}{\sigma^2_{\log n_{\mathrm{g}}}}, \tag{14}$$

where $n_{\mathrm{g}}$ is a number density of LBGs in the subsample. We calculate the number density of LBGs corrected for incompleteness using the UV luminosity functions of Ono et al. (2017) and Bouwens et al. (2015). The LBG number densities are presented in Table 4. We assume 10% fractional uncertainties in the number densities as Zheng et al. (2007). This 10% uncertainty is a conservative assumption, because the actual uncertainty is less than 5% (Ono et al. 2017). We constrain the parameters of our HOD model using the Markov Chain Monte Carlo (MCMC) parameter estimation technique.

In our HOD model, an occupation function for central galaxies follows a step function with a smooth transition,

$$\langle N_{\mathrm{cen}}(M_{\mathrm{h}})\rangle = \frac{1}{2}\left[1 + \mathrm{erf}\left(\frac{\log M_{\mathrm{h}} - \log M_{\mathrm{min}}}{\sqrt{2}\sigma_{\log M_{\mathrm{h}}}}\right)\right]. \qquad (15)$$

An occupation function for satellite galaxies is expressed by a power law with a mass cut,

$$\langle N_{\mathrm{sat}}(M_{\mathrm{h}})\rangle = \langle N_{\mathrm{cen}}(M_{\mathrm{h}})\rangle\left(\frac{M_{\mathrm{h}} - M_{\mathrm{cut}}}{M_{\mathrm{sat}}}\right)^{\alpha}. \qquad (16)$$

The total occupation function is

$$\langle N_{\mathrm{tot}}(M_{\mathrm{h}})\rangle = \langle N_{\mathrm{cen}}(M_{\mathrm{h}})\rangle + \langle N_{\mathrm{sat}}(M_{\mathrm{h}})\rangle. \qquad (17)$$

These functional forms are motivated by N-body simulations, smoothed particle hydrodynamic simulations, and semi-analytic models for low-$z$ galaxies and LBGs (e.g., Kravtsov et al. 2004; Zheng et al. 2005; Garel et al. 2015).

These occupation functions assume a mass complete sample, while our sample is constructed with the LBG selection methods. Since LBG selection methods require galaxies to be detected in the rest-frame UV band with not too-red colors, we may miss dusty starburst galaxies at the targeted redshift. For example, our $z \sim 4$ color criteria (Equations (1)-(3)) cannot select dusty galaxies with $E(B - V) > 0.6$. We have investigated the effect of this missing population on the HOD modeling using the COSMOS 2015 catalog (Laigle et al. 2016). We calculate fractions of dusty galaxies with $E(B - V) > 0.6$ in the stellar-mass threshold subsamples at $z_{\mathrm{phot}} = 3.3 - 4.2$. The fractions are 2% and 19% for our faintest ($m_{\mathrm{UV}} < 25.5$) and brightest ($m_{\mathrm{UV}} < 24.0$) HSC subsamples with the stellar mass thresholds of $\log M_{\star}^{\mathrm{th}} = 9.41$ and 10.62, respectively (see Section 4.3.1 for the stellar mass thresholds). As presented in Table 4, the effective galaxy biases estimated in our study are $4.02^{+0.05}_{-0.08}$ and $6.66^{+0.09}_{-0.07}$ for the faintest and brightest subsamples, respectively (see Table 4). If we assume that the galaxy bias of these missing 2% and 19% population is $b_{\mathrm{g}} = 7$ (Webb et al. 2003; Weiß et al. 2009), expected galaxy biases are 4.06 and 6.77, respectively, within $1\sigma$ errors of our estimates. In addition, we investigate the effect of sources not detected in rest-frame optical bands and missed in the photo-$z$ catalog. Even if we assume that the submillimeter galaxies not detected in the $K$-band in Simpson et al. (2017) are all $z \sim 4$ sources, the fractions are small, 1% and 8% for the faintest and brightest subsamples, respectively. Thus we have concluded that the effect of the dusty starburst galaxies is not significant.

We calculate the mean dark matter halo mass of central and satellite galaxies, $\langle M_{\mathrm{h}}\rangle$, effective galaxy bias, $b_{\mathrm{g}}^{\mathrm{eff}}$, and the satellite fraction, $f_{\mathrm{sat}}$, as follows:

$$\langle M_{\mathrm{h}}\rangle = \frac{1}{n_{\mathrm{g}}}\int dM_{\mathrm{h}}\frac{dn}{dM_{\mathrm{h}}}(M_{\mathrm{h}},z)N_{\mathrm{tot}}(M_{\mathrm{h}})M_{\mathrm{h}}, \qquad (18)$$

$$b_{\mathrm{g}}^{\mathrm{eff}} = \frac{1}{n_{\mathrm{g}}}\int dM_{\mathrm{h}}\frac{dn}{dM_{\mathrm{h}}}(M_{\mathrm{h}},z)N_{\mathrm{tot}}(M_{\mathrm{h}})b_{\mathrm{h}}(M_{\mathrm{h}},z), \qquad (19)$$

$$f_{\mathrm{sat}} = \frac{1}{n_{\mathrm{g}}}\int dM_{\mathrm{h}}\frac{dn}{dM_{\mathrm{h}}}(M_{\mathrm{h}},z)N_{\mathrm{sat}}(M_{\mathrm{h}}), \qquad (20)$$

where $\frac{dn}{dM_{\mathrm{h}}}(M_{\mathrm{h}},z)$, $b_{\mathrm{h}}(M_{\mathrm{h}},z)$, and $n_{\mathrm{g}}$ are the halo func-tion, halo bias, and the galaxy number density in the model (Equation (51) in Harikane et al. 2016), respectively. We assume the Behroozi et al. (2013b) halo mass function, the NFW dark matter halo profile (Navarro et al. 1996, 1997), the Duffy et al. (2008) concentration parameter, and the Smith et al. (2003) non-linear matter power spectrum.

Since some theoretical studies claim that the halo bias is scale dependent in the quasi-linear scale of $r \sim 50$ Mpc (the non-linear halo bias effect; Reed et al. 2009; Jose et al. 2013, 2016, 2017), we implement two models for the halo bias. In the first case (called "linear HOD model"), we assume the scale-independent linear halo bias of Tinker et al. (2010), $b_{\mathrm{lin}}(M_{\mathrm{h}},z)$. The ACFs at $10'' < \theta < 90''$ are not used in this case, because they could be affected by the non-linear halo bias effect. In the second case (called "non-linear HOD model"), we assume the scale-dependent non-linear halo bias of Jose et al. (2016),

$$b_{\mathrm{nl}}(r,M_{\mathrm{h}},z) = b_{\mathrm{lin}}(M_{\mathrm{h}},z)\zeta(r,M_{\mathrm{h}},z), \qquad (21)$$

where $\zeta(r,M_{\mathrm{h}},z)$ is the scale-dependent correction factor. In Jose et al. (2016), $\zeta(r,M_{\mathrm{h}},z)$ is expressed as a function of the peak height, $\nu(M_{\mathrm{h}},z) = \delta_{\mathrm{c}}/\sigma(M_{\mathrm{h}},z)$, where $\delta_{\mathrm{c}} = 1.69$ and $\sigma(M_{\mathrm{h}},z)$ are the critical linear over-density and the variance of matter fluctuation on a mass scale $M_{\mathrm{h}}$ at the redshift of $z$, respectively. We assume $\zeta = \zeta(\nu = 2)$ ($\zeta(\nu = 3)$) for $\nu > 2$ (3) at $z < 4.5$ ($z > 4.5$), because we find that the ACFs are overpre-dicted by a factor of $\sim 2 - 10$ at $10'' < \theta < 90''$ without these constraints compared to the ACFs at large scale ($\theta \sim 1000''$). $\zeta$ is extrapolated in the scale smaller than $r = 3$ Mpc where the fit-ting function is not calibrated in Jose et al. (2016). In this case, all of the ACFs are used for the fitting. As detailed in Section 4.1, we find that our results are largely unchanged, regardless of which HOD model we use.

Both HOD models have 5 parameters, $M_{\mathrm{min}}$, $\sigma_{\log M_{\mathrm{h}}}$, $M_{\mathrm{cut}}$, $M_{\mathrm{sat}}$, and $\alpha$. In our previous work (Harikane et al. 2016), we also fit the duty cycle. However, we find that we can reproduce the observed ACFs and number densities without the duty cy-cle in Section 4.1. In addition, Harikane et al. (2016) use the fitting formula of the halo mass function presented in Tinker et al. (2010) without normalization constraints (see Appendix C in Tinker et al. 2008), which overestimates the abundance by a factor of $\sim 1.7$. As a result, the estimate in Harikane et al. (2016) is more consistent with a duty cycle of 1. Thus, we as-sume that the duty cycle is unity in this work. We take $M_{\mathrm{min}}$ and $M_{\mathrm{sat}}$ as free parameters, which control typical masses of halos having one central and satellite galaxies, respectively. We fix $\sigma_{\log M_{\mathrm{h}}} = 0.2$ and $\alpha = 1.0$, following results of previous stud-ies (e.g., Kravtsov et al. 2004; Zheng et al. 2005; Conroy et al. 2006; Ishikawa et al. 2016b). To derive $M_{\mathrm{cut}}$ from $M_{\mathrm{min}}$, we use the relation

$$M_{\mathrm{cut}} = M_{\mathrm{min}}^{-0.5}, \qquad (22)$$

which is given by Coupon et al. (2015). Because the exact value of $M_{cut}$ has very little importance compared to the other parameters, this assumption does not change any of our conclusions. We use Equation (55) in Harikane et al. (2016) for fitting of $z \sim 6$ subsample, because we cannot obtain a good constraint on $M_{sat}$. For comparison, we re-calculate the best-fit HOD parameters for ACFs and number densities of subsamples constructed from the Hubble data in Harikane et al. (2016) in the same cosmological parameter sets and halo mass definition as this work.

# 4 Results

## 4.1 Results of the HOD fittings

We plot the observed ACFs and their best-fit models by the linear HOD models in Figure 12. The best-fit parameters and their $1\sigma$ errors are presented in Table 4. The linear HOD models can reproduce the observed ACFs at small ($\theta < 10''$) and large ($\theta > 90''$) scales. The agreement in the small scale implies that the assumption for satellite galaxies in the HOD model is valid at $z \gtrsim 4$; the spatial distribution of satellite galaxies follows the density profile of the dark matter halo.

However, the linear HOD models underpredict the ACFs by a factor of $1.5 - 3$ in $10'' < \theta < 90''$, the transition scale between 1- and 2-halo terms (the quasi-linear scale), except for the $z \sim 4$ $m_{UV}^{th} = 25.0$, and $z \sim 5$ $m_{UV}^{th} = 24.0, 24.5$ subsamples. These results indicate that the ACFs at $10'' < \theta < 90''$ can not be explained by the scale-independent halo bias due to the non-linear halo bias effect. Actually, the reduced $\chi^2$ values increase if we include the ACF data in $10'' < \theta < 90''$ for the calculation of Equation (14), except for the $z \sim 4$ $m_{UV}^{th} = 25.0$, and $z \sim 5$ $m_{UV}^{th} = 24.0, 24.5$ subsamples whose ACFs are well explained by the linear HOD model. These excesses in the observed ACFs cannot be explained by changing the concentration parameter to very small value, e.g., $c = 0.1$.

The best-fit ACFs by the non-linear HOD models are presented in Figure 13. Table 5 summarizes the best-fit parameters and their $1\sigma$ errors. The amplitudes of the ACFs on the quasi-linear scale increase compared with those by the linear HOD models. The non-linear HOD models can reproduce the ACFs of $z \sim 4$ $m_{UV}^{th} = 25.5$, and $z \sim 5$ $m_{UV}^{th} = 25.0, 25.5$ subsamples with reducing the $\chi^2$ values, but are still not enough to explain the ACFs of $z \sim 4$ $m_{UV} < 24.0, 24.5$ subsamples. Note that Jose et al. (2016), from which we take our assumed relation for the scale-dependent non-linear halo bias, use the halo mass defined by the friends-of-friends algorithm, while we define the halo mass by the spherical overdensity, $M_{200}$. It is known that projected correlation functions of the halos defined by the spherical overdensity and the friends-of-friends algorithm differ by a factor of $\sim 2$ at separations of $\sim 1$ Mpc (Reid & Spergel 2009). This difference could explain the discrepancies between the best-fit non-linear HOD models and our observations. In seven out of the nine subsamples, the best-fit parameters in the non-linear HOD models agree with those in the linear HOD models within the $1\sigma$ errors, because the amplitude of the 2-halo term is determined by the data at $\theta > 90''$ in the linear HOD model, where the correction factor $\zeta$ in the non-linear HOD model is almost unity. Thus we adopt the linear HOD model as our fiducial model, and the following results and discussions are based on it. Note that our conclusions do not change if we adopt the non-linear HOD model.

We compare the observed number densities and predictions from the HOD model in Figure 14. The predictions agree well with the observed ones, indicating that the unity duty cycle can explain both the number densities and the ACFs.

## 4.2 $M_{UV} - M_h$ relation

Figure 15 shows the results of the halo mass, $M_h$, as a function of the UV magnitude, $M_{UV}$. The measurements are summarized in Table 6. We plot $M_{min}$ and $M_{UV}^{th}$ as $M_h$ and $M_{UV}$, respectively, for a fair comparison with abundance matching studies. The halo mass of LBGs found in the HSC data ranges from $6 \times 10^{11}$ $M_\odot$ to $4 \times 10^{12}$ $M_\odot$, which is more massive than those of LBGs in the Hubble data (Harikane et al. 2016). The combination of the Hubble and HSC data allows us to investigate the $M_{UV} - M_h$ relation over 2 orders of magnitude in the halo mass at $z \sim 4$ and 5. There is a positive correlation between the UV luminosity and the halo mass, indicating that more UV-luminous galaxies reside in more massive halos (see also Park et al. 2016, 2017). The slope of the $M_{UV} - M_h$ relation becomes steeper at the brighter magnitude, which is similar to the local $M_* - M_h$ relation (e.g., Leauthaud et al. 2012; Coupon et al. 2015).

We find a redshift evolution of the $M_{UV} - M_h$ relation from $z \sim 4$ to $z \sim 6(7)$ at the $> 10(5.2)\sigma$ confidence level at $M_{UV} = -21.7(-19.3)$ mag. At given $M_{UV}$, $M_h$ monotonically decreases from $z \sim 4$ to $z \sim 6(7)$ by $\sim 0.4$ dex at $M_{UV} = -21.7(-19.3)$ mag. In other words, the dust-uncorrected star formation rate (SFR) increases with redshift at fixed dark matter halo mass. We also plot the $M_{UV} - M_h$ relations at $z \sim 4$, 5, 6, and 7 in Figure 15 as the blue, green, orange, and red solid curves, respectively. These relations are expressed with the following equation:

$$\log_{10}(f_{M_{UV}-M_h}^{-1}) = \log_{10}(M_h(M_{UV}))$$
$$= \log M_1 + \beta \left[ -\frac{1}{2.5}(M_{UV} - M_{UV,0}) \right]$$
$$+ \frac{10^{-\frac{1}{2.5}(M_{UV}-M_{UV,0})\delta}}{1 + 10^{-\frac{1}{2.5}(M_{UV}-M_{UV,0})(-\gamma)}} - \frac{1}{2} \quad (23)$$

where $M_1$ ($M_{UV,0}$) is characteristic halo mass (UV magnitude). $\beta$, $\delta$, and $\gamma$ control the low-mass slope, high-mass slope, and curvature of the relation, respectively. If we assume a power-law relation between the UV magnitude

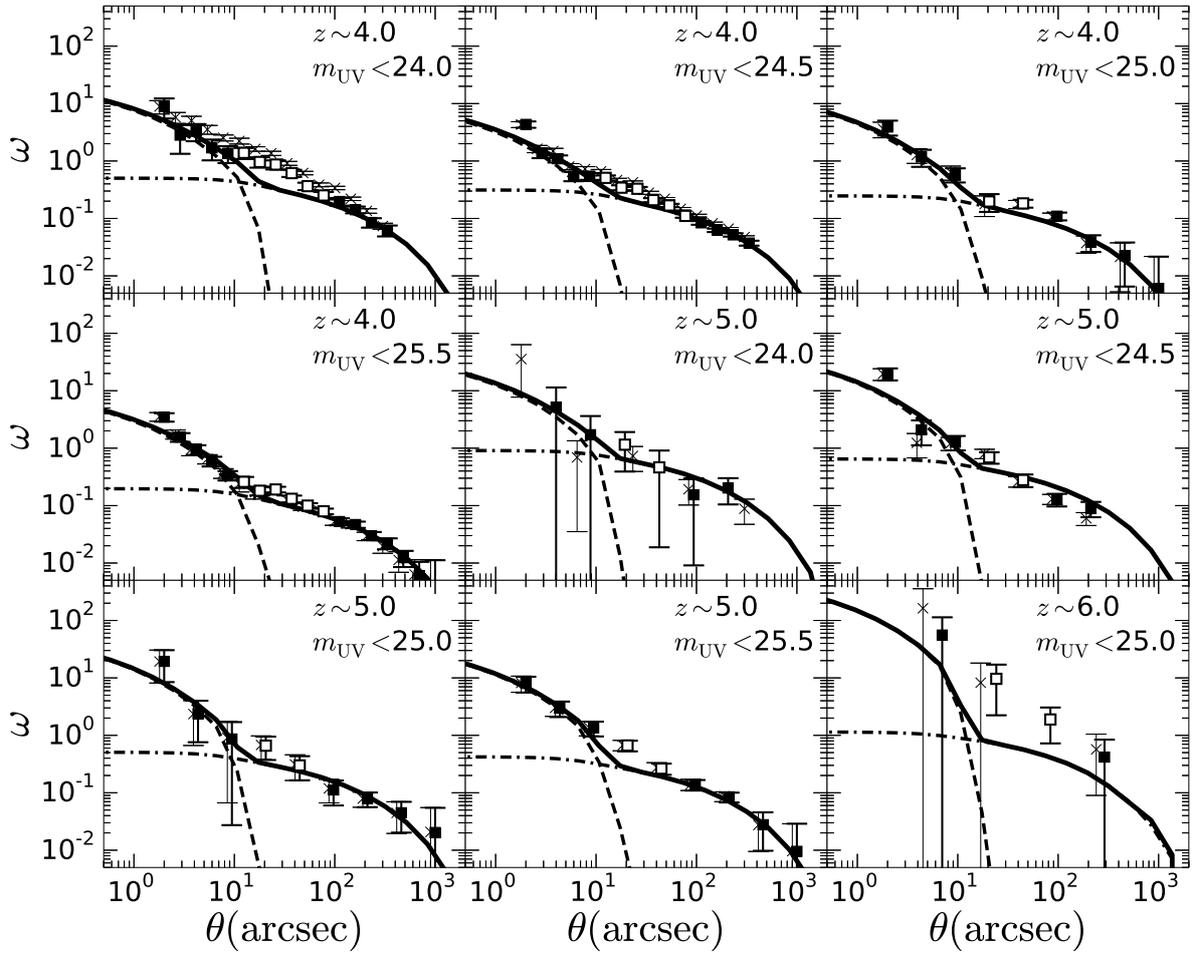

**Fig. 12.** Observed ACFs of the subsamples and their best-fit by the linear HOD model. The black squares denote the observed ACFs of the subsamples at each redshift. The ACF data presented with the open squares in $10'' < \theta < 90''$ are not used in our HOD fitting because they are affected by the non-linear halo bias effect (see Section 3.3). The dashed and dot-dashed curves represent the one-halo and two-halo terms, respectively, and the solid curves are the summations of the one-halo and two-halo terms. For reference, we also plot ACFs of the subsamples with the magnitude cut of $m_{\rm UV}^{\rm cut} = 20.0$ mag instead of $23.5$ mag with the crosses. The crosses are shifted by $0.05$ dex for clarity.

**Table 4.** Summary of the Clustering Measurements with the Linear HOD Model

| z | $m_{\rm UV}^{\rm th}$ | N | $n_{\rm obs}$ | $M_{\rm UV}^{\rm th}$ | $\log SFR$ | $\log M_\star^{\rm th}$ | $\log M_{\rm min}$ † | $\log M_{\rm sat}$ | $\log \langle M_{\rm h} \rangle$ | $b_{\rm g}^{\rm eff}$ | $\log f_{\rm sat}$ | $\chi^2/{\rm dof}$ |
|---|---|---|---|---|---|---|---|---|---|---|---|---|
| (1) | (2) | (3) | (4) | (5) | (6) | (7) | (8) | (9) | (10) | (11) | (12) | (13) |
| 3.8 | 24.0 | 16,392 | $1.1 \times 10^{-5}$ | $-22.0$ | 2.11 (1.59) | 10.62 | $12.61^{+0.02}_{-0.02}$ | $14.88^{+0.17}_{-0.12}$ | $12.63^{+0.01}_{-0.02}$ | $6.66^{+0.09}_{-0.07}$ | $-2.29^{+0.10}_{-0.17}$ | 0.51 (1.16) |
| 3.8 | 24.5 | 77,656 | $6.9 \times 10^{-5}$ | $-21.5$ | 1.87 (1.39) | 10.19 | $12.22^{+0.02}_{-0.02}$ | $14.23^{+0.08}_{-0.09}$ | $12.32^{+0.03}_{-0.02}$ | $5.29^{+0.07}_{-0.07}$ | $-2.06^{+0.06}_{-0.06}$ | 3.38 (3.85) |
| 3.8 | 25.0 | 3,138 | $2.3 \times 10^{-4}$ | $-21.0$ | 1.62 (1.19) | 9.79 | $11.94^{+0.03}_{-0.03}$ | $13.26^{+0.81}_{-0.09}$ | $12.12^{+0.02}_{-0.02}$ | $4.54^{+0.05}_{-0.05}$ | $-1.54^{+0.09}_{-0.69}$ | 0.58 (0.56) |
| 3.8 | 25.5 | 8,083 | $6.0 \times 10^{-4}$ | $-20.5$ | 1.38 (0.99) | 9.41 | $11.69^{+0.03}_{-0.02}$ | $12.51^{+0.13}_{-0.07}$ | $11.98^{+0.01}_{-0.02}$ | $4.02^{+0.05}_{-0.10}$ | $-1.32^{+0.04}_{-0.16}$ | 0.76 (2.66) |
| 4.9 | 24.0 | 3,108 | $4.0 \times 10^{-6}$ | $-21.0$ | 2.34 (1.77) | 10.49 | $12.49^{+0.02}_{-0.02}$ | $15.08^{+1.57}_{-0.39}$ | $12.43^{+0.03}_{-0.05}$ | $8.47^{+0.08}_{-0.09}$ | $-2.73^{+0.37}_{-1.59}$ | 0.53 (0.44) |
| 4.9 | 24.5 | 14,083 | $2.2 \times 10^{-6}$ | $-21.9$ | 2.09 (1.57) | 10.16 | $12.21^{+0.01}_{-0.02}$ | $14.15^{+0.31}_{-0.09}$ | $12.21^{+0.01}_{-0.02}$ | $7.14^{+0.17}_{-0.06}$ | $-2.14^{+0.08}_{-0.30}$ | 2.95 (2.42) |
| 4.9 | 25.0 | 758 | $7.2 \times 10^{-5}$ | $-21.4$ | 1.83 (1.37) | 9.84 | $11.95^{+0.02}_{-0.02}$ | $13.28^{+0.41}_{-0.17}$ | $12.02^{+0.02}_{-0.02}$ | $6.21^{+0.09}_{-0.06}$ | $-1.77^{+0.17}_{-0.39}$ | 0.29 (0.59) |
| 4.9 | 25.5 | 2,064 | $1.9 \times 10^{-4}$ | $-20.9$ | 1.58 (1.17) | 9.53 | $11.74^{+0.02}_{-0.02}$ | $12.43^{+0.15}_{-0.13}$ | $11.89^{+0.01}_{-0.12}$ | $5.59^{+0.05}_{-0.05}$ | $-1.46^{+0.05}_{-0.12}$ | 0.29 (1.30) |
| 5.9 | 25.0 | 411 | $8.6 \times 10^{-6}$ | $-21.7$ | 2.02 (1.49) | 9.74 | $12.04^{+0.01}_{-0.02}$ | (12.80) | $12.02^{+0.02}_{-0.02}$ | $8.88^{+0.13}_{-0.04}$ | $(-1.84)$ | 0.66 (0.81) |

Columns: (1) Mean redshift. (2) Threshold apparent magnitude in the rest-frame UV band. (3) Number of galaxies in the subsample. (4) Number density of galaxies in the subsample in units of $\mathrm{Mpc}^{-3}$. (5) Threshold absolute magnitude in the rest-frame UV band. (6) SFR corresponding to $M_{\rm UV}^{\rm th}$ after the extinction correction in units of $M_\odot \, \mathrm{yr}^{-1}$. The value in the parenthesis is the SFR before the correction. (7) Threshold stellar mass estimated in this study. (8) Best-fit value of $M_{\rm min}$ in units of $M_\odot$. (9) Best-fit value of $M_{\rm sat}$ in units of $M_\odot$. The value in parenthesis is derived from $M_{\rm min}$ via Equation (55) in Harikane et al. (2016). (10) Mean halo mass in units of $M_\odot$. (11) Effective bias. (12) Satellite fraction. (13) Reduced $\chi^2$ value. The values in the parentheses are the reduced $\chi^2$ values when the ACFs in all scales are used.
† The best-fit values for the magnitude cut of $m_{\rm UV}^{\rm cut} = 20.0$ mag instead of 23.5 mag are $\log M_{\rm min} = [12.64^{+0.02}_{-0.01}, 12.23^{+0.02}_{-0.03}, 11.95^{+0.02}_{-0.02}, 11.73^{+0.02}_{-0.03}]$, for $z \sim 4$ $m_{\rm UV}^{\rm th} = [24.0, 24.5, 25.0, 25.5]$, $\log M_{\rm min} = [12.49^{+0.02}_{-0.02}, 12.19^{+0.02}_{-0.01}, 11.93^{+0.02}_{-0.02}, 11.71^{+0.03}_{-0.01}]$ for $z \sim 5$ $m_{\rm UV}^{\rm th} = [24.0, 24.5, 25.0, 25.5]$, and $\log M_{\rm min} = 12.02^{+0.04}_{-0.02}$ for $z \sim 6$ $m_{\rm UV}^{\rm th} = 25.0$.

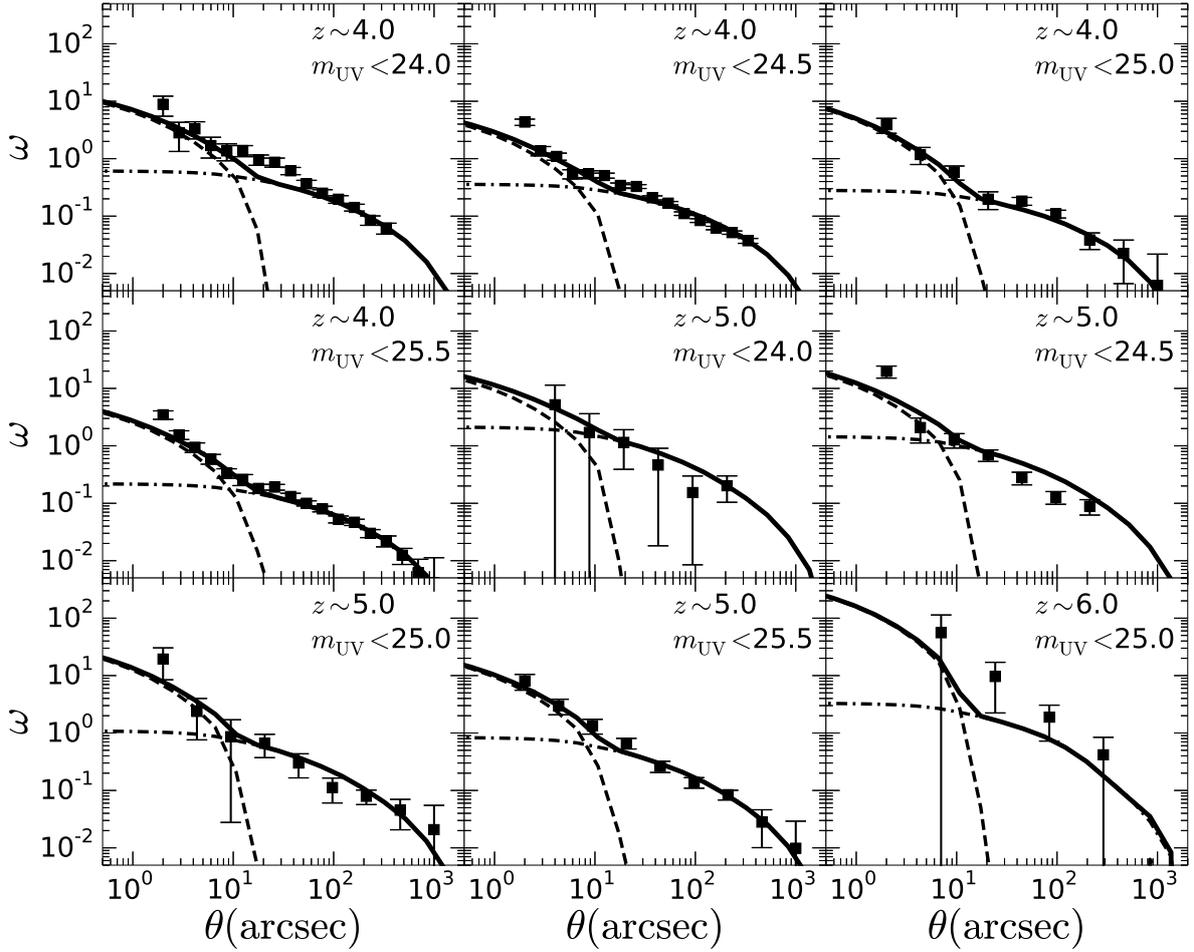

**Fig. 13.** Same as Figure 12 but with non-linear HOD model.

**Table 5.** Summary of the Clustering Measurements with the Non-Linear HOD Model

| $z$ | $m_{\rm UV}^{\rm th}$ | $\log M_{\rm min}$ | $\log M_{\rm sat}$ | $\log \langle M_{\rm h} \rangle$ | $b_{\rm g}^{\rm eff}$ | $\log f_{\rm sat}$ | $\chi^2/{\rm dof}$ |
|---|---|---|---|---|---|---|---|
| (1) | (2) | (3) | (4) | (5) | (6) | (7) | (8) |
| 3.8 | 24.0 | $12.61^{+0.03}_{-0.01}$ | $14.95^{+0.14}_{-0.14}$ | $12.63^{+0.01}_{-0.02}$ | $6.68^{+0.06}_{-0.10}$ | $-2.36^{+0.11}_{-0.15}$ | 0.97 |
| 3.8 | 24.5 | $12.22^{+0.02}_{-0.03}$ | $14.33^{+0.08}_{-0.10}$ | $12.33^{+0.01}_{-0.02}$ | $5.32^{+0.05}_{-0.10}$ | $-2.15^{+0.09}_{-0.05}$ | 4.59 |
| 3.8 | 25.0 | $11.95^{+0.01}_{-0.04}$ | $13.27^{+0.12}_{-0.13}$ | $12.13^{+0.00}_{-0.03}$ | $4.55^{+0.03}_{-0.10}$ | $-1.54^{+0.08}_{-0.16}$ | 0.72 |
| 3.8 | 25.5 | $11.70^{+0.02}_{-0.03}$ | $12.61^{+0.13}_{-0.10}$ | $11.97^{+0.01}_{-0.02}$ | $4.02^{+0.04}_{-0.06}$ | $-1.40^{+0.05}_{-0.10}$ | 0.55 |
| 4.9 | 24.0 | $12.42^{+0.01}_{-0.02}$ | $15.40^{+0.50}_{-0.22}$ | $12.38^{+0.01}_{-0.01}$ | $8.11^{+0.08}_{-0.08}$ | $-3.22^{+0.24}_{-0.43}$ | 0.85 |
| 4.9 | 24.5 | $12.08^{+0.01}_{-0.02}$ | $14.80^{+0.60}_{-0.47}$ | $12.12^{+0.00}_{-0.02}$ | $6.63^{+0.01}_{-0.12}$ | $-2.63^{+0.12}_{-0.92}$ | 4.16 |
| 4.9 | 25.0 | $11.95^{+0.02}_{-0.02}$ | $13.33^{+0.43}_{-0.17}$ | $12.02^{+0.01}_{-0.02}$ | $6.20^{+0.07}_{-0.07}$ | $-1.81^{+0.17}_{-0.39}$ | 0.38 |
| 4.9 | 25.5 | $11.74^{+0.03}_{-0.02}$ | $12.51^{+0.28}_{-0.10}$ | $11.88^{+0.01}_{-0.03}$ | $5.57^{+0.08}_{-0.08}$ | $-1.54^{+0.09}_{-0.22}$ | 0.53 |
| 5.9 | 25.0 | $12.04^{+0.01}_{-0.02}$ | $(12.80)$ | $12.02^{+0.02}_{-0.01}$ | $8.88^{+0.13}_{-0.04}$ | $(-1.84)$ | 0.77 |

Columns: (1) Mean redshift. (2) Threshold apparent magnitude in the rest-frame UV band. (3) Best-fit value of $M_{\rm min}$ in units of $M_\odot$. (4) Best-fit value of $M_{\rm sat}$ in units of $M_\odot$. The value in the parenthesis is derived from $M_{\rm min}$ via Equation (55) in Harikane et al. (2016). (5) Mean halo mass in units of $M_\odot$. (6) Effective bias. (7) Satellite fraction. (8) Reduced $\chi^2$ value.

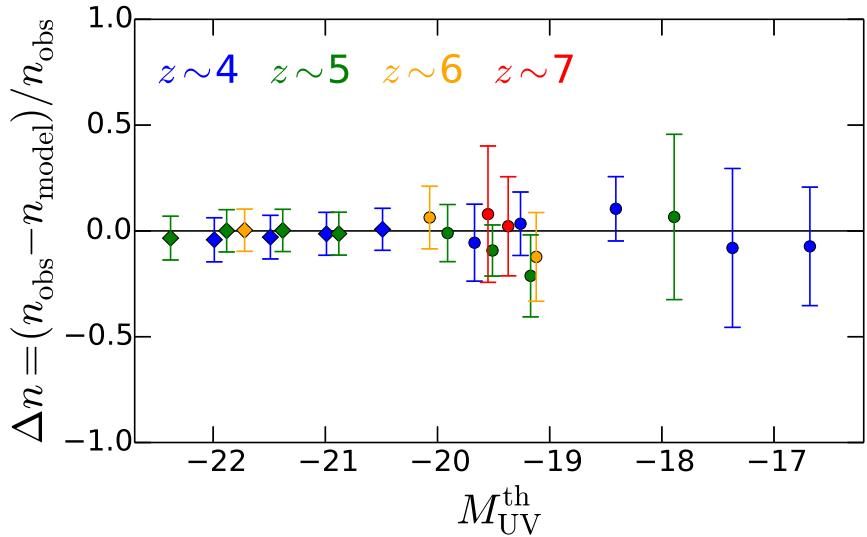

**Fig. 14.** Comparison of the number densities between the linear HOD models and observations. The blue, green, orange, and red squares (circles) represent the relative differences of the number densities between the HOD models and observations for the subsamples in this work (Harikane et al. 2016), at $z \sim 4, 5, 6$, and $7$, respectively, as a function of the threshold absolute UV magnitudes, $M_{\rm UV}^{\rm th}$.

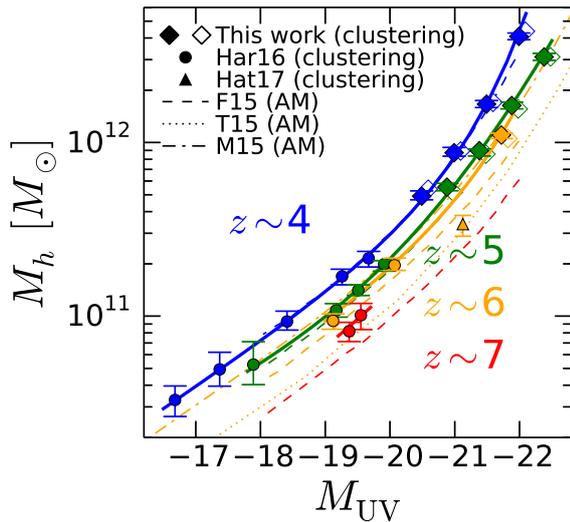

**Fig. 15.** $M_{\rm UV} - M_{\rm h}$ relation. The blue, green, orange, and red filled diamonds (circles) denote the halo masses as a function of the UV magnitude, at $z \sim 4, 5, 6$, and $7$, respectively, for the subsamples in this work (Harikane et al. 2016, Har16). Note that the halo masses shown by the circles are slightly different from original values in Harikane et al. (2016), because we re-calculate the halo masses with the latest Planck cosmology and different halo mass definition from Harikane et al. (2016). The blue, green, orange, and red solid curves show the relations of Equation (23) at $z \sim 4, 5, 6$, and $7$, respectively. The orange triangle denotes the result of Hatfield et al. (2017, Hat16). The dashed, dotted, and dot-dashed curves are the results of the abundance matching studies by Finkelstein et al. (2015, F15), Trac et al. (2015, T15), and Mason et al. (2015, M15), respectively. For reference, we also plot results of the subsamples with the magnitude cut of $m_{\rm UV}^{\rm cut} = 20.0$ mag with the open diamonds. The open diamonds are shifted by $0.1$ mag for clarity.

and the stellar mass, Equation (23) is the same as the one proposed by Behroozi et al. (2010). We use parameter sets of $(M_{\rm UV,0}, \log M_1, \beta, \delta, \gamma) = (-21.49, 12.22, 0.63, 0.84, 0.30)$ for $z \sim 4$, $(-20.88, 11.70, 0.47, 0.26, 0.68)$ for $z \sim 5$, $(-21.89, 12.16, 0.63, 0.74, 0.30)$ for $z \sim 6$, and $(-21.49, 11.89, 0.63, 0.74, 0.30)$ for $z \sim 7$. Note that these solid curves are just for illustrative purposes. We calculate the significances of the redshift evolution based on not these solid curves, but the individual data points.

We compare the halo masses of our results with those of the clustering study of Hatfield et al. (2017), and abundance matching studies of Finkelstein et al. (2015), Trac et al. (2015), and Mason et al. (2015) in Figure 15. Although the halo mass estimate in Hatfield et al. (2017) is lower than an interpolation of our results, their estimate is consistent with ours at the $\sim 2\sigma$ level given its uncertainty. We find that our clustering measurements are in good agreement with those of the abundance matching studies at $z \sim 4$ and 5. At $z \sim 6$, our halo mass estimates are consistent with the abundance matching results of Finkelstein et al. (2015) and Mason et al. (2015) at the $1 - 2\sigma$ level, but larger than the curve of Trac et al. (2015) at the $\sim 3\sigma$ level. These good agreement may be due to low satellite fractions in the high redshift universe as discussed in Harikane et al. (2016).

In Figure 16, we compare the mean halo masses, $\langle M_{\rm h} \rangle$, of our subsamples with the literature. Because most of the previous studies assume $(\Omega_{\rm m}, \Omega_{\Lambda}, h, \sigma_8) = (0.3, 0.7, 0.7, 0.9)$ that is different from our assumption, we obtain HOD model fitting results for our data with $(\Omega_{\rm m}, \Omega_{\Lambda}, h, \sigma_8) = (0.3, 0.7, 0.7, 0.9)$ for comparison. Similarly, the results of the previous studies are re-calculated with the same cosmological parameter sets. In

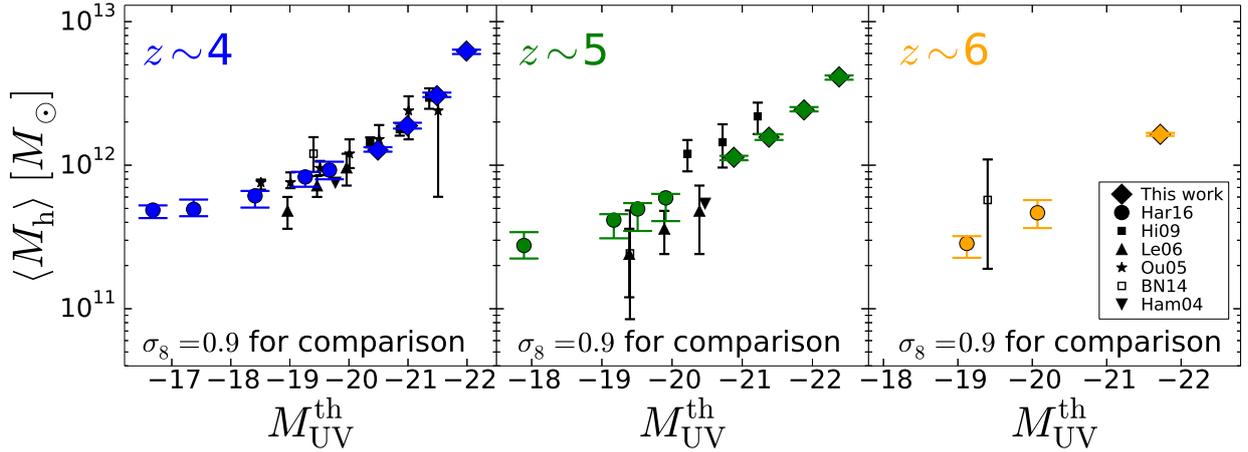

**Fig. 16.** Comparison of the mean dark matter halo masses with the literature under the same cosmology. The diamonds represent the mean dark matter halo masses in this work with the cosmological parameters of $(h, \Omega_m, \Omega_L, \sigma_8) = (0.7, 0.3, 0.7, 0.9)$. The circles are the results of Harikane et al. (2016, Har16) with the same cosmology. The black symbols denote the results of the previous studies. We plot the results of Hamana et al. (2004, Ham04; downward triangle), Ouchi et al. (2005, Ou05; stars), Lee et al. (2006, Le06; upward triangles) and Hildebrandt et al. (2009, Hi09; diamonds). The downward triangles have no error bars, because Hamana et al. (2004) do not provide errors of the mean dark matter halo mass. We also show the results of Barone-Nugent et al. (2014, BN14) as black open squares, who use the simple power law model. We compile the results of Barone-Nugent et al. (2014) in the same cosmology.

**Table 6.** Halo Mass, SHMR, and $SFR/\dot{M}_h$ in This Study.

| $z$ | $M_{UV}^{th}$ | $\log M_h$ | $M_*/M_h \ (10^{-3})$ | $SFR/\dot{M}_h \ (10^{-2})$ |
|-----|---------------|------------|-----------------------|------------------------------|
| (1) | (2) | (3) | (4) | (5) |
| 3.8 | $-22.0$ | $12.61^{+0.02}_{-0.02}$ | $9.70^{+0.43}_{-0.44}$ | $1.18^{+0.05}_{-0.06}$ |
| 3.8 | $-21.5$ | $12.22^{+0.02}_{-0.02}$ | $8.87^{+0.41}_{-0.41}$ | $1.79^{+0.09}_{-0.09}$ |
| 3.8 | $-21.0$ | $11.94^{+0.03}_{-0.02}$ | $6.65^{+0.45}_{-0.31}$ | $2.05^{+0.14}_{-0.10}$ |
| 3.8 | $-20.5$ | $11.69^{+0.03}_{-0.02}$ | $4.98^{+0.34}_{-0.24}$ | $2.18^{+0.11}_{-0.17}$ |
| 3.8 | $-19.6$ | $11.33^{+0.04}_{-0.05}$ | $3.24^{+0.28}_{-0.40}$ | $2.12^{+0.25}_{-0.22}$ |
| 3.8 | $-19.2$ | $11.23^{+0.04}_{-0.05}$ | $2.35^{+0.21}_{-0.28}$ | $1.73^{+0.20}_{-0.18}$ |
| 3.8 | $-18.4$ | $10.97^{+0.06}_{-0.05}$ | $1.48^{+0.19}_{-0.18}$ | $1.27^{+0.16}_{-0.20}$ |
| 3.8 | $-17.3$ | $10.69^{+0.10}_{-0.10}$ | $0.85^{+0.17}_{-0.15}$ | $0.78^{+0.16}_{-0.21}$ |
| 3.8 | $-16.7$ | $10.52^{+0.08}_{-0.10}$ | $0.59^{+0.10}_{-0.11}$ | $0.55^{+0.11}_{-0.12}$ |
| 4.9 | $-22.4$ | $12.49^{+0.02}_{-0.02}$ | $9.01^{+0.41}_{-0.42}$ | $1.75^{+0.09}_{-0.09}$ |
| 4.9 | $-21.9$ | $12.21^{+0.02}_{-0.02}$ | $8.03^{+0.36}_{-0.36}$ | $1.94^{+0.09}_{-0.10}$ |
| 4.9 | $-21.4$ | $11.95^{+0.01}_{-0.02}$ | $7.02^{+0.16}_{-0.51}$ | $2.04^{+0.05}_{-0.15}$ |
| 4.9 | $-20.9$ | $11.74^{+0.02}_{-0.02}$ | $5.65^{+0.27}_{-0.33}$ | $1.89^{+0.09}_{-0.10}$ |
| 4.9 | $-19.9$ | $11.30^{+0.02}_{-0.04}$ | $4.39^{+0.21}_{-0.43}$ | $1.72^{+0.16}_{-0.09}$ |
| 4.9 | $-19.5$ | $11.15^{+0.03}_{-0.03}$ | $3.82^{+0.26}_{-0.28}$ | $1.55^{+0.11}_{-0.12}$ |
| 4.9 | $-19.1$ | $11.03^{+0.04}_{-0.04}$ | $3.35^{+0.29}_{-0.32}$ | $1.38^{+0.13}_{-0.14}$ |
| 4.9 | $-17.9$ | $10.72^{+0.13}_{-0.12}$ | $1.68^{+0.44}_{-0.51}$ | $0.65^{+0.16}_{-0.25}$ |
| 5.9 | $-21.7$ | $12.04^{+0.01}_{-0.02}$ | $7.39^{+0.18}_{-0.35}$ | $1.74^{+0.08}_{-0.05}$ |
| 5.9 | $-20.0$ | $11.29^{+0.04}_{-0.03}$ | $6.22^{+0.56}_{-0.48}$ | $1.29^{+0.09}_{-0.13}$ |
| 5.9 | $-19.1$ | $10.97^{+0.04}_{-0.04}$ | $4.35^{+0.38}_{-0.38}$ | $0.82^{+0.09}_{-0.08}$ |
| 6.8 | $-19.5$ | $11.00^{+0.07}_{-0.08}$ | $4.51^{+0.66}_{-0.94}$ | $0.88^{+0.16}_{-0.16}$ |
| 6.8 | $-19.3$ | $10.91^{+0.05}_{-0.06}$ | $4.53^{+0.51}_{-0.66}$ | $0.87^{+0.12}_{-0.11}$ |

Columns: (1) Mean redshift. (2) Threshold absolute magnitude in the rest-frame UV band. (3) Dark matter halo mass ($M_{min}$) in units of $M_\odot$. (4) SHMR in units of $10^{-3}$. (5) Ratio of the SFR to the dark matter accretion rate in units of $10^{-2}$.

this way, we conduct our comparisons using an equivalent set of cosmological parameters across all data sets. In Figure 16, we find that our $z \sim 4$ results are consistent with those of the previous studies within the uncertainties. While the previous results at $z \sim 5$ are largely scattered, our $z \sim 5$ results are placed near the center of the distribution of the previous studies. At $z \sim 6$, our results agree with that of Barone-Nugent et al. (2014). In summary, our results are consistent with most of the previous studies. Furthermore, our results improve on both the statistics and the dynamic range covered in $M_{UV}$.

## 4.3 SHMR

### 4.3.1 Stellar Mass Estimate

We estimate stellar masses of our LBGs to derive SHMRs. Since $M_{UV} - M_*$ relations at high redshift are constrained with recent Hubble and *Spitzer* observations, some studies (Shibuya et al. 2015, 2016; Harikane et al. 2016; Ishikawa et al. 2016b) use the $M_{UV} - M_*$ (or $SFR - M_*$) relations to estimate stellar masses. In Figure 17, we plot recent estimates of the $M_{UV} - M_*$ relations from the literature. Stellar mass estimates from Shibuya et al. (2015) are higher than those of Salmon et al. (2015) and Song et al. (2016), because Shibuya et al. (2015) use the SED-fitting results in Skelton et al. (2014) without nebula emission. Ishikawa et al. (2016b) use a $SFR - M_*$ relation presented in Tanaka (2015). We convert the SFR to the observed UV magnitude, $M_{UV}$, using an attenuation-UV slope ($\beta_{UV}$) relation (Meurer et al. 1999) and $\beta_{UV} - M_{UV}$ relations (Bouwens et al. 2014), and plot the converted relation in Figure 17. We find that Ishikawa et al. (2016b) overestimate stellar masses by $0.5 - 1.0$ dex compared to those of Song et al. (2016).

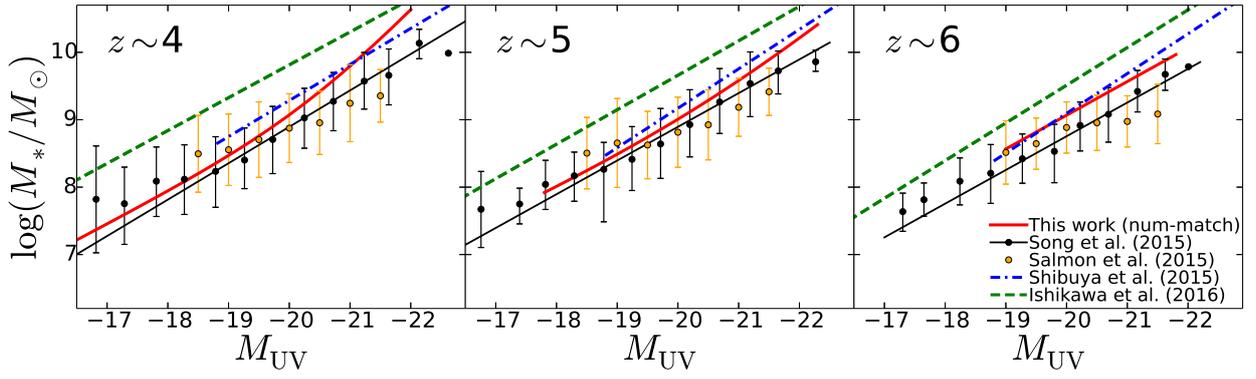

**Fig. 17.** $M_{UV} - M_*$ relation used in this study. The red solid curves represent the $M_{UV} - M_*$ relations used in this study (the num-match relation; see Section 4.3.1 for details), which are corrected for the difference between the $M_{UV}$-threshold and $M_*$-threshold samples to some extent. The black line and circles are the results from Song et al. (2016), while orange circles are from Salmon et al. (2015). The blue dot-dashed and green dashed lines are the relations in Shibuya et al. (2015) and Ishikawa et al. (2016b), respectively. The stellar mass is in the Chabrier (2003) IMF.

Our following estimates are based on the latest results in Song et al. (2016), which are derived from the SED-fitting with nebula emission with the deep Hubble and *Spitzer* data.

Since our sample is not $M_*$-threshold sample like low-redshift studies but $M_{UV}$-threshold one, we need to be careful in the stellar mass estimate, especially for bright galaxies. We explain a systematic bias by using a simple $M_{UV} - M_*$ relation. Let us consider a $M_{UV} - M_*$ relation defined by median values of $M_*$ in given $M_{UV}$ bins, $M_*^{median}(M_{UV})$ (the black solid line in Figure 18). If the scatter in the $M_{UV} - M_*$ relation is negligible, the $M_*^{median}(M_{UV})$-threshold sample is identical to the $M_{UV}$-threshold sample. However, with a scatter, the number of galaxies in the $M_*^{median}(M_{UV})$-threshold sample (N=587 in Figure 18) should be larger than that of the $M_{UV}$-threshold sample (N=299), since abundant UV-faint galaxies are scattered into the $M_*^{median}(M_{UV})$-threshold sample. As a result, the clustering strength of the $M_*^{median}(M_{UV})$-threshold sample would be weaker than that of the $M_{UV}$-threshold sample. This effect becomes more significant in the bright end with steeper slope of the UV luminosity function. Thus we cannot use $M_*^{median}(M_{UV})$ directly as the stellar mass of the bright $M_{UV}$-threshold sample in deriving the SHMR, because the halo mass of $M_*^{median}(M_{UV})$-threshold sample would be quite different from that of the $M_{UV}$-threshold sample estimated in Section 4.2.

We can reduce this systematic uncertainty by keeping the numbers of galaxies same in both $M_{UV}$ and $M_*$-threshold samples (the red-dashed line in Figure 18). In Figure 17, we plot the $M_*$-threshold value of a sample which contains the same number of galaxies as a given $M_{UV}$-threshold sample (the red-solid curve; the num-match relation), assuming the $M_{UV} - M_*$ relation of Song et al. (2016) with the scatter and the UV luminosity function of Ono et al. (2017). Our num-match relation is close to the original $M_*^{median}(M_{UV})$ relation (the black solid line) at the faint magnitude ($-19 < M_{UV} < -17$), and deviates from the original relation at the bright magnitude ($M_{UV} < -21$) due to the steep slope in the luminosity function. Still, our num-match relation is located between the other $M_{UV} - M_*$ relations in the magnitude range we are interested in. We estimate the threshold stellar mass, $M_*^{th}$, from $M_{UV}^{th}$ using this num-match relation at each redshift (Table 4). Note that this num-match relation cannot completely avoid the systematic uncertainty. $M_*$-threshold subsamples are needed for further discussion. To understand uncertainties of $z \sim 7$ estimates, we derive two stellar masses for each $z \sim 7$ subsample using two num-match relations based on $z \sim 6$ and 7 $M_{UV} - M_*$ relations in Song et al. (2016).

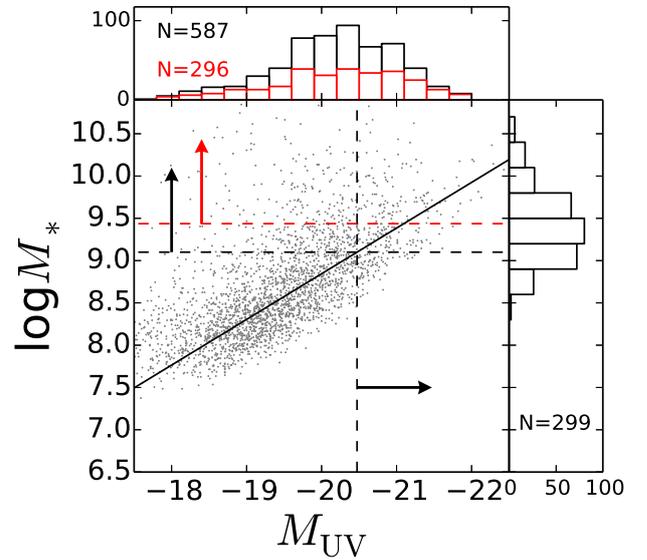

**Fig. 18.** Stellar mass estimate and its systematic uncertainty. The gray dots (black-solid line) are the stellar mass for individual galaxies (a median relation) in Song et al. (2016). The vertical black-dashed line represents the UV magnitude threshold of $M_{UV} < -20.4$ mag. The horizontal black-dashed line is the corresponding $M_*^{median}(M_{UV})$ threshold of $\log(M_*/M_\odot) > 9.1$. The number of galaxies in the $M_*^{median}(M_{UV})$-threshold sample is larger than that of the $M_{UV}$-threshold sample. The horizontal red-dashed line denotes that stellar mass threshold of $\log(M_*/M_\odot) > 9.4$ in which number of galaxies is comparable to that of $M_{UV}$-threshold sample.

### 4.3.2 SHMRs and the Evolution

Figure 19 and Table 6 show the results of SHMRs at each redshift. We plot $M_{\min}$ and $M_*^{th}/M_{\min}$ as $M_h$ and $M_*/M_h$, respectively. We find that the SHMR ranges from $\sim 10^{-3}$ to $\sim 10^{-2}$, and increases with increasing $M_h$ at each redshift. Moreover, the slope of the SHMR becomes flatter at more massive halos at $z \sim 4$, and 5. This flattening indicates the AGN feedback and/or inefficient gas cooling regulating the star formation in the massive halos, which will be discussed in Section 5.1.

We compare our SHMRs with those of previous studies using clustering analysis and abundance matching methods. Ishikawa et al. (2016b) estimate SHMRs by clustering analysis at $z \sim 3 - 5$ using the CFHTLS data. The SHMRs in Ishikawa et al. (2016b) are higher than ours at $z \sim 4$ and 5, because Ishikawa et al. (2016b) overestimate the stellar mass compared to ours (see Figure 17). Note that we plot the original values in Ishikawa et al. (2016b) rather than corrected values based on Figure 17, because actual UV magnitudes of their subsamples are not provided in Ishikawa et al. (2016b). Our SHMRs at $z \sim 4$ are comparable to those at $z \sim 3$ in Durkalec et al. (2017), who use clustering analysis with $\sim 3000$ spectroscopic galaxies. Stefanon et al. (2016) estimate the SHMRs at $z \sim 4 - 7$ by the abundance matching using rest-frame optical galaxy luminosity functions. We find that their estimates agree with ours at $z \sim 4$ within their $1\sigma$ error, and are slightly higher than ours at $\sim 2\sigma$ levels at $z \sim 5$ and 6. Our estimates are lower than those of the abundance matching study of Finkelstein et al. (2015), because they use the median stellar mass to estimate the SHMR instead of the threshold stellar mass. Our SHMRs agree well with those of Moster et al. (2013) at $z \sim 4$ within their uncertainty ($\sim 0.4$ dex in the SHMR). SHMRs of Behroozi et al. (2013b) are comparable with ours at $z \sim 4$ and 5 in the massive halo of $M_h > 2 \times 10^{12} M_\odot$. Their estimates are higher than ours in the mass range of $M_h < 2 \times 10^{12} M_\odot$ by a factor of $\sim 2$, but still within their $2\sigma$ errors (see also; Sun & Furlanetto 2016; Rodríguez-Puebla et al. 2017).

To discuss the redshift evolution, we present our SHMRs at $z \sim 4 - 7$ in Figure 20, with that at $z \sim 0$ in Behroozi et al. (2013b) for comparison. We plot two cases for the $z \sim 7$ SHMRs using the $z \sim 7$ and $z \sim 6$ num-match relations, to understand the uncertainty in the stellar mass estimate at $z \sim 7$. We also show the SHMRs at fixed halo masses of $10^{11}$ and $10^{12} M_\odot$ as functions of the redshift in Figure 21 with literature. For studies without stellar mass estimates, we calculate stellar masses from UV luminosity limits using relations in Shibuya et al. (2015). At $M_h \sim 1 \times 10^{11} M_\odot$, the SHMR decreases from $z \sim 0$ to 4 by a factor of $\sim 4$, and increases from $z \sim 4$ to 7 by a factor of $\sim 4$, despite the uncertainties in the $z \sim 7$ stellar mass estimates. The SHMR at $M_h \sim 1 \times 10^{12} M_\odot$ also decreases from $z \sim 0$ to 4 by a factor of $\sim 3$, but does not evolve significantly from $z \sim 4$ to 6, similar to the abundance matching result of Stefanon et al. (2016). The decreasing trend of the SHMR from $z \sim 0$ to 4 is consistent with previous weak lensing and clustering studies (the lower panel in Figure 21). These redshift evolutions cannot be explained by the missing dusty star burst population in our LBG sample. As discussed in Section 3.3, if we consider the effect of the dusty star burst galaxies, the expected effective halo bias (and the halo mass) still agrees with our estimate within the $1\sigma$ error. We will compare these SHMR redshift evolutions with theoretical studies in Section 5.2.

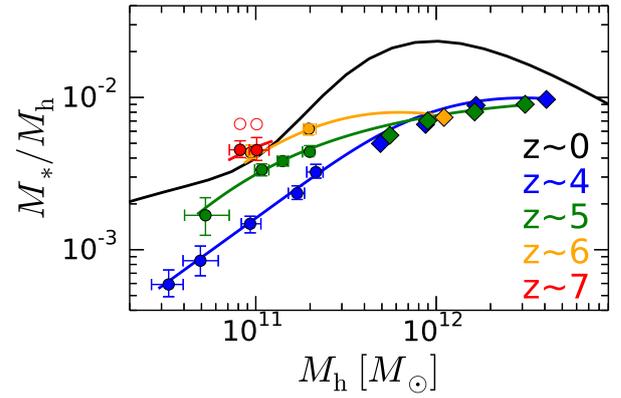

**Fig. 20.** SHMR evolution with redshift. The blue, green, orange, and red diamonds (circles) are the results in this work (Harikane et al. 2016), and the curves represent Equation (23), at $z \sim 4, 5, 6,$ and 7, respectively. The statistical errors for our data are smaller than the symbols (diamonds). The filled red circles denote the SHMRs from the $z \sim 7$ num-match relation, while the open red circles are from the $z \sim 6$ relation, as Harikane et al. (2016). The black curve represents the SHMR of Behroozi et al. (2013b) at $z \sim 0$, which is computed with the same cosmological parameters and halo mass definition as in our analysis. The uncertainties of the SHMR of Behroozi et al. (2013b) are $0.01$ and $0.008$ dex at $M_h = 10^{11}$ and $10^{12} M_\odot$, respectively.

### 4.4 $SFR/\dot{M}_h - M_h$ Relation at $z \sim 4 - 7$

We estimate $SFR/\dot{M}_h$ which is a ratio of the SFR to the dark matter accretion rate. We derive the dust-uncorrected SFR ($SFR_{uncorr}$) from $M_{UV}^{th}$ using the following calibration (Kennicutt 1998):

$$SFR_{uncorr} \, (M_\odot \, yr^{-1}) = 1.4 \times 10^{-28} L_{UV} \, (erg \, s^{-1} \, Hz^{-1}). \quad (24)$$

Note that the accuracy of this calibration is typically $\sim 15\%$ (Calzetti 2013). We correct the SFR for the dust extinction ($SFR_{corr}$) using an attenuation-UV slope ($\beta_{UV}$) relation (Meurer et al. 1999) and $\beta_{UV} - M_{UV}$ relations (Bouwens et al. 2014). The estimated SFRs are presented in Table 4. We calculate $\dot{M}_h$ as a function of halo mass and redshift using an analytic formula obtained from N-body simulation results

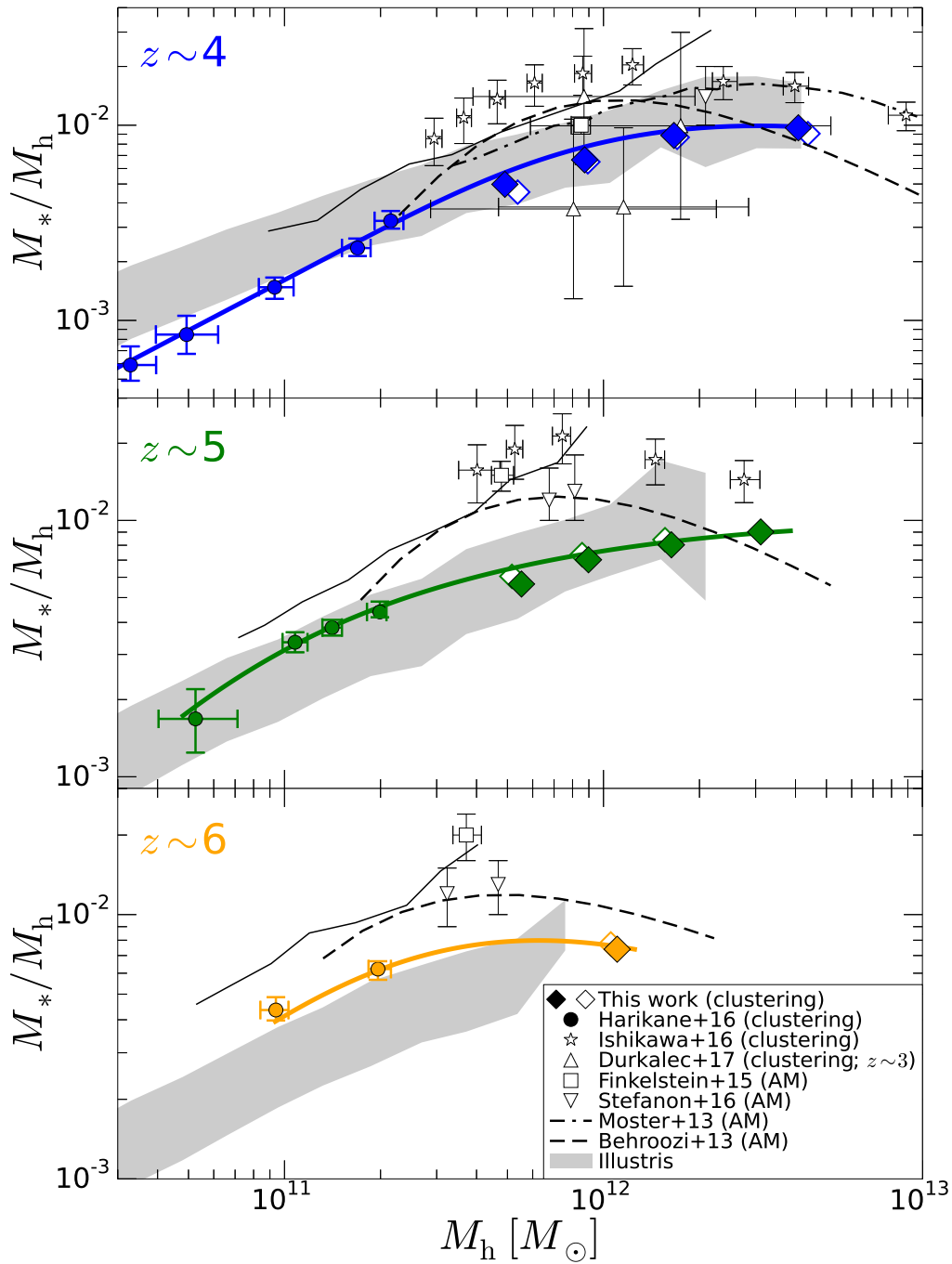

**Fig. 19.** SHMR at each redshift. The filled diamonds (circles) are the SHMRs for the subsamples in this work (Harikane et al. 2016). The statistical errors for our data are smaller than the symbols (diamonds). Note that the positions of the circles are not the same as Harikane et al. (2016), because the cosmological parameters, halo mass definition, and stellar mass estimates are different from Harikane et al. (2016). The solid curves show relations of Equation (23) with our stellar mass estimates. The stars and upward triangles are the results from clustering analyses by Ishikawa et al. (2016b) and Durkalec et al. (2017), respectively. The squares, downward triangles, dot-dashed curve, and dashed curves denote the SHMRs by the abundance matching studies of Finkelstein et al. (2015), Stefanon et al. (2016), Moster et al. (2013) and, Behroozi et al. (2013b), respectively. The dashed curves are different from original ones in Behroozi et al. (2013b), as these curves are recomputed with the same cosmological parameters and halo mass definition as ours. The predictions from the illustris simulation are shown with the gray shaded regions. For reference, we also plot results of the subsamples with the magnitude cut of $m_{\mathrm{UV}}^{\mathrm{cut}} = 20.0$ mag with the open diamonds.

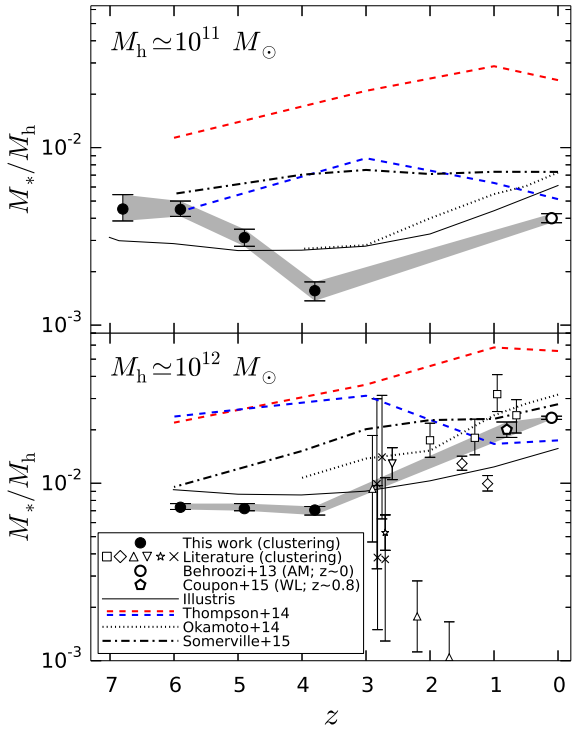

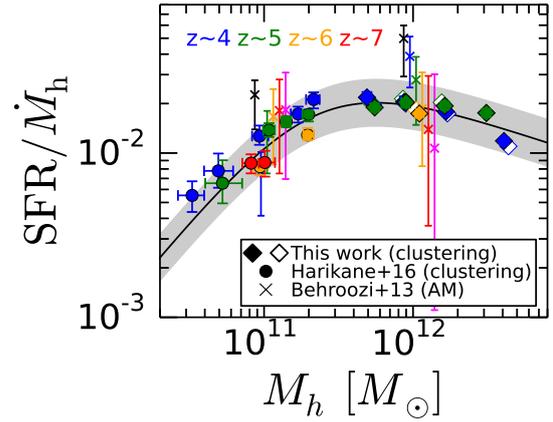

**Fig. 22.** $SFR/\dot{M}_{\rm h}$ as a function of the halo mass. The filled blue, green, orange, and red circles (circles) denote the ratio of the extinction-corrected SFR to the dark matter accretion rate at $z \sim 4, 5, 6$, and 7, respectively, in this work (Harikane et al. 2016). The statistical errors for our data are smaller than the symbols (diamonds). We also plot $SFR/\dot{M}_{\rm h}$ at $z \sim 0, 4, 5, 6, 7$, and 8 from Behroozi et al. (2013b) with the black, blue, green, orange, red, and magenta crosses for comparison. The black solid curve is the fitting formulae of Equation (25) for the $SFR/\dot{M}_{\rm h} - M_{\rm h}$ relation for $z \sim 4 - 7$, and the gray shaded region represents the typical scatter (0.15 dex) of the data points. For reference, we also plot results of the subsamples with the magnitude cut of $m_{\rm UV}^{\rm cut} = 20.0$ mag with the open diamonds.

**Fig. 21.** Upper panel: SHMRs at the fixed dark matter halo mass of $M_{\rm h} = 10^{11} M_\odot$. The filled circles represent the observed SHMRs in this study at $z \sim 4 - 7$, and the open circle is the observational result at $z \sim 0$ (Behroozi et al. 2013b). The dashed, dotted, dot-dashed, and solid lines denote the predictions from theoretical studies of Thompson et al. (2014), Okamoto et al. (2014), Somerville et al. (2015), and Illustris, respectively. The red and blue dashed lines correspond to the Fiducial and $H_2$ models in Thompson et al. (2014). The gray shaded region shows the possible redshift evolution. Lower panel: same as the upper panel but for $M_{\rm h} = 10^{12} M_\odot$. We also plot the results from literature; McCracken et al. (2015), Wake et al. (2011, diamonds), Adelberger et al. (2005, upward triangles), Bielby et al. (2013, downward triangles), Trainor & Steidel (2012, star), Durkalec et al. (2017, crosses), and Coupon et al. (2015, pentagon).

(Behroozi et al. 2013b).

We plot the ratio of $SFR_{\rm corr}/\dot{M}_{\rm h}$ at $z \sim 4 - 7$ as a function of the halo mass in Figure 22. The measurements are also summarized in Table 6. The black solid curve in Figure 22 represents the following $SFR/\dot{M}_{\rm h} - M_{\rm h}$ relation:

$$\frac{SFR}{\dot{M}_{\rm h}} = \frac{2 \times 1.7 \times 10^{-2}}{(M_{\rm h}/10^{11.35})^{-1.1} + (M_{\rm h}/10^{11.35})^{0.3}}. \quad (25)$$

Interestingly, we do not find any significant redshift evolution of $SFR/\dot{M}_{\rm h}$ beyond 0.15 dex (the gray shade in Figure 22) at $z \gtrsim 4$. Behroozi et al. (2013a) discuss that the ratio of the SFR to the baryon accretion rate change very little at $z = 0 - 4$ by their abundance matching. Bian et al. (2013) also report similar redshift independence at $z \sim 3 - 5$ by clustering. We confirm it with the large and homogeneous sample at $z \sim 4 - 7$ by the clustering analysis in the wide dynamical range of $4 \times 10^{10} M_\odot < M_{\rm h} < 4 \times 10^{12} M_\odot$. We will discuss the implications of this fundamental (redshift-independent)

$SFR/\dot{M}_{\rm h} - M_{\rm h}$ relation in Section 5.4.

We also plot the ratios of $SFR_{\rm corr}/\dot{M}_{\rm h}$ from Behroozi et al. (2013b) in Figure 22. The ratios of Behroozi et al. (2013b) at $z \sim 0$ are systematically higher than ours at $z \sim 4 - 7$ by a factor of $\sim 3$, although the errors are large. If we take this possible evolution to $z \sim 0$ into account, the ratio can be expressed as

$$\frac{SFR}{\dot{M}_{\rm h}} = \frac{2 \times 5.1 \times 10^{-2} \times (1+z)^{-0.6}}{(M_{\rm h}/10^{11.35})^{-1.1} + (M_{\rm h}/10^{11.35})^{0.3}}. \quad (26)$$

With this equation, the ratio still does not change beyond 0.15 dex at $z \sim 4 - 7$.

### 4.5 Satellite fraction

We plot the estimated satellite fractions as a function of the stellar mass threshold, $M_*^{\rm th}$, in Figure 23. The best-fit satellite fraction ranges from $2 \times 10^{-3}$ to $8 \times 10^{-2}$ at $z \sim 4 - 5$, and the subsamples with high $M_*^{\rm th}$ tend to have lower satellite fractions. The satellite fraction at $z \sim 6$ in Hatfield et al. (2017) is lower than ours at $z \sim 4 - 5$. In addition, the satellite fractions of the $z \sim 5$ subsamples are tentatively smaller than those of $z \sim 4$ ones at fixed stellar mass. Our results at $z \sim 4 - 5$ are in good agreement with those of Ishikawa et al. (2016b) at $M_*^{\rm th} \lesssim 10^{10} M_\odot$, but lower than Ishikawa et al. (2016b) at $M_*^{\rm th} \gtrsim 10^{10} M_\odot$, probably due to the overestimate of the stellar mass in Ishikawa et al. (2016b) compared to ours (see Figure 17). Our satellite fractions are lower than the predictions from the illustris simulation (Vogelsberger et al. 2014; Genel et al. 2014), which is a suite of cosmological hydrodynamical simulations.

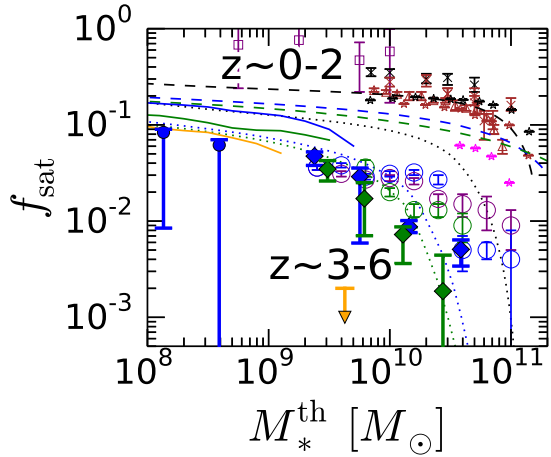

**Fig. 23.** Satellite fraction as a function of the stellar mass threshold $M_*^{\rm th}$. The blue and green filled symbols show the satellite fractions in this work at $z \sim 4$ and $5$, respectively. The orange downward arrow shows the upper limit of the satellite fraction at $z \sim 6$ in Hatfield et al. (2017). The purple, blue, and green open circles are the results of Ishikawa et al. (2016b) at $z \sim 3$, $4$, and $5$, respectively. The purple open squares represent the satellite fractions in Durkalec et al. (2017). For comparison, we plot the $z \sim 0 - 2.5$ satellite fractions of Wake et al. (2011, crosses), Martinez-Manso et al. (2015, open upward triangles), and McCracken et al. (2015, stars) with black ($0.0 < z < 1.2$), brown ($1.2 < z < 2.0$), and magenta ($2.0 < z < 2.5$) colors. The blue, green, and orange solid curves are the predictions from the illustris simulation. The dashed and dotted curves are the calculated subhalo fractions of Equation (35) and (36), respectively, at $z \sim 0$ (black), $4$ (blue), and $5$ (green).

We compare the satellite fractions at $z \sim 4 - 5$ with those of $0 < z < 2.5$ galaxies. The satellite fractions at $z \sim 0 - 2$ are typically $\sim 0.1$ in $10^{10} M_\odot < M_*^{\rm th} < 10^{11} M_\odot$, and do not evolve strongly from $z \sim 0$ to $2$. Our estimates at $z \sim 4 - 5$ are smaller than those of $z \sim 0 - 2$ galaxies by about one order of magnitude at fixed $M_*^{\rm th}$. Satellite fractions of $2 < z < 2.5$ galaxies in McCracken et al. (2015) are $0.02 - 0.08$, located between those of $z \sim 0 - 2$ and $z \sim 4 - 5$. We will discuss interpretations of these results in Section 5.3.

## 5 Discussion

### 5.1 Feedback Effects at High Redshift

We discuss the feedback effects based on our observational results using a simple analytic model (the $t_{\rm cool}$ model). In the $t_{\rm cool}$ model, the SFR is proportional to a ratio of all gas mass, $M_{\rm gas}$, to a cooling time scale, $t_{\rm cool}$,:

$$SFR = \epsilon_{\rm SF} \frac{M_{\rm gas}}{t_{\rm cool}}, \tag{27}$$

where $\epsilon_{\rm SF}$ represents star formation efficiency. We assume $\epsilon_{\rm SF} = 0.1$ and $M_{\rm gas} = f_{\rm b} M_{\rm h}$ where $f_{\rm b} = \Omega_{\rm b}/\Omega_{\rm m} = 0.159$ is the cosmic baryon fraction. The cooling time scale is a ratio of the thermal energy per unit volume, $E_{\rm thermal}$, to the cooling rate, $|\dot{E}_{\rm cool}|$,

$$t_{\rm cool} = \frac{E_{\rm thermal}}{|\dot{E}_{\rm cool}|}. \tag{28}$$

The thermal energy is derived from the gas number density, $n$, and the temperature, $T$,

$$E_{\rm thermal} = \frac{3}{2} n k_{\rm B} T, \tag{29}$$

where $k_{\rm B}$ is the Boltzmann constant. For $T$, we use the virial temperature,

$$T = \frac{\mu}{2 k_{\rm B}} \frac{GM}{r_{200}}, \tag{30}$$

with $\mu = 1.4$. The cooling rate is expressed as

$$|\dot{E}_{\rm cool}| = n^2 \Lambda(T, Z), \tag{31}$$

where $\Lambda(T, Z)$ is a cooling function depending on the temperature, $T$, and the metallicity, $Z$. We use the cooling function of Sutherland & Dopita (1993). We set the metallicity which is inferred from the mass-metallicity relation at $z \sim 3.5$ presented in Maiolino et al. (2008), allowing for its $1\sigma$ scatter.

In Figure 24, we compare the model-estimated $SFR/M_{\rm h}$ with our observational results. In the lower mass range of $M_{\rm h} \lesssim 10^{12} M_\odot$, the observational increasing trends toward higher halo mass are not reproduced by the $t_{\rm cool}$ model at $z \sim 4, 5$. This indicates that the gas cooling effect alone cannot explain the observed SFR in the low-mass halos. We need to consider a negative feedback effect which suppresses star formation more efficiently in lower mass halos. The SNe feedback is one of the candidates, because the outflow gas would escape from the halo more easily in the lower mass halo with the shallower gravitational potential.

In the higher mass range of $M_{\rm h} \gtrsim 10^{12} M_\odot$, the slope of model-estimated $SFR/M_{\rm h}$ is comparable or steeper than the observational results at both $z \sim 4$ and $5$, although the uncertainties are large. This indicates that the observed decreasing (flat) $SFR/M_{\rm h}$ in the high-mass halos at $z \sim 4(5)$ can be explained by the decrease of the gas cooling efficiency with higher virial temperature. Note that these results do not exclude the possibility of the AGN feedback which may be efficient in massive halos. Although the AGN feedback could steepen the slope of the $SFR/M_{\rm h}$-$M_{\rm h}$ relation, the merger-driven starburst or another feedback may flatten the slope. In addition, it is not clear whether our model reproduces the observed absolute values of $SFR/M_{\rm h}$, because the star formation efficiency is assumed to be $\epsilon = 0.1$.

We also compare the observed $SFR/M_{\rm h}$ values with those of the Illustris simulation. We find that the increasing trend of $SFR/M_{\rm h}$ with $M_{\rm h}$ up to $M_{\rm h} \sim 10^{12} M_\odot$ agrees with our observational results. Note that the discussions above do not change if we consider the 15% uncertainty on Equation (24).

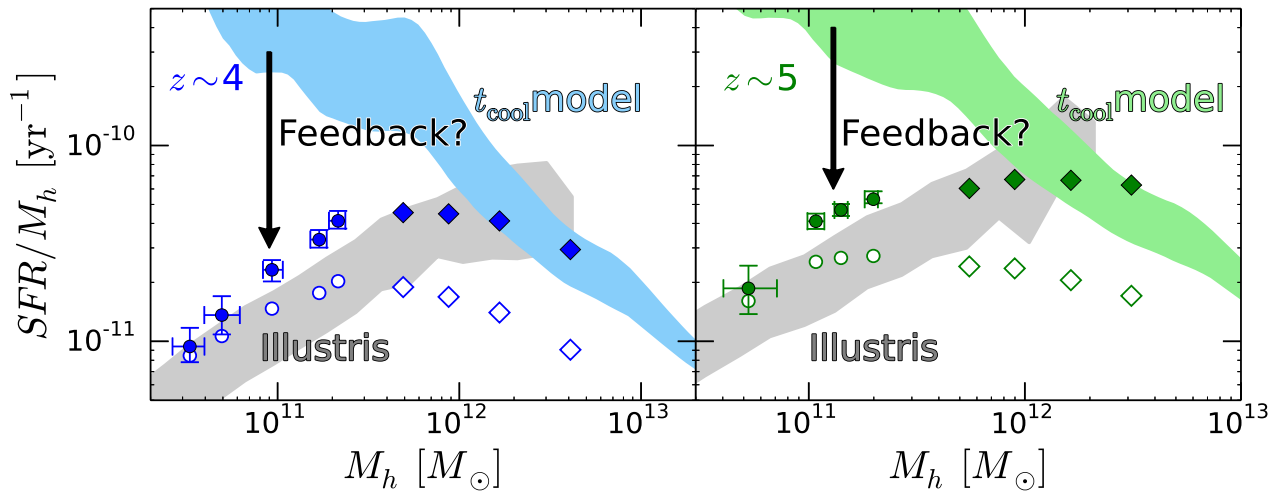

**Fig. 24.** $SFR/M_h$ as a function of the halo mass. The filled (open) symbols show the ratio of the extinction-corrected (uncorrected) SFR to the dark matter halo mass. The statistical errors for our data are smaller than the symbols (diamonds). The cyan and light-green shaded regions represent the predictions by the $t_{cool}$ model at $z = 4$ and $5$, respectively (see Section 5.1 for details). The predictions from the Illustris simulation are plotted with the gray shaded region.

## 5.2 Comparison with Theoretical Models

We find that the observed SHMR at $M_h \simeq 10^{11}\ M_\odot$ decreases by a factor of $\sim 4$ from $z \sim 0$ to $z \sim 4$, and increases by a factor of $\sim 4$ from $z \sim 4$ to $z \sim 7$. Also, the SHMR at $M_h \simeq 10^{12}\ M_\odot$ decreases by a factor of $\sim 3$ from $z \sim 0$ to $z \sim 4$, but does not significantly evolve from $z \sim 4$ to $z \sim 6$. We compare these observed SHMR results with previous theoretical studies in Figure 21. Thompson et al. (2014) predict SHMRs at $z = 0 - 6$ by cosmological hydrodynamic simulations including the effects of the SN feedback and reionization, without and with effect of molecular hydrogen (the Fiducial and $H_2$ models in Thompson et al. 2014, respectively). Okamoto et al. (2014) also use cosmological hydrodynamic simulations including SN, radiation pressure, and AGN feedbacks, and predict SHMRs at $z = 0 - 4$. Somerville et al. (2015) predict SHMRs using semi-analytic models with SN and AGN feedbacks at $z \sim 0 - 6$. We compare the GK model of Somerville et al. (2015) with our results. The Illustris simulation considers various physical processes including the SNe and AGN feedbacks (Vogelsberger et al. 2014; Genel et al. 2014). In Figure 21, we plot the predictions of the SHMRs from these models as functions of the redshift at the fixed halo mass of $M_h = 10^{11}$ and $10^{12}\ M_\odot$.

We compare the evolutional trends of the SHMRs by ignoring the normalizations. The decreases of the SHMRs from $z \sim 0$ to $z \sim 4$ are reproduced by some models (i.e., the Fiducial model in Thompson et al. 2014; Okamoto et al. 2014; Somerville et al. 2015, Illustris) at both $M_h = 10^{11}$ and $10^{12}\ M_\odot$. Although the Illustris simulation tentatively shows little increase of SHMR from $z \sim 4$ to $7$, no models compared here can reproduce the observed increasing trend at $M_h = 10^{11}\ M_\odot$, as discussed in Harikane et al. (2016). At $M_h = 10^{12}\ M_\odot$, the observed SHMR is almost constant at $z \sim 4 - 6$, similar to that in the Illustris simulation. However, models other than Illustris show SHMRs tentatively decreasing from $z \sim 4$ to $z \sim 6$. These comparisons indicate that the star formation efficiency would be underestimated at high redshift in the theoretical models compared here, especially at $M_h = 10^{11}\ M_\odot$. Note that the recent BLUETIDES simulation reproduces the increasing trend of the SHMR at $z > 7$ (Waters et al. 2016; Bhowmick et al. 2017).

## 5.3 Interpretations of Very Low Satellite Fractions

The estimated satellite fractions at $z \sim 4 - 5$ are smaller than those at $z \sim 0 - 2$ by about one order of magnitude (Figure 23). Three possibilities are considered to explain this huge discrepancy. The first is the difference in the sample selections. For example, the galaxies in the $z \sim 0 - 2$ samples are stellar-mass-selected samples, while $z \sim 4 - 5$ galaxies are the UV-selected LBGs not including passive galaxies. If the passive galaxies have more satellite galaxies than the star-forming galaxies like LBGs, this discrepancy could be explained. Although Coupon et al. (2012) report that satellite fractions of full and red (passive) galaxies are very similar; the difference is less than a factor of 1.5 at $z < 1.2$, we cannot eliminate the possibility of very different satellite fractions between passive and star-forming galaxies at $z \gtrsim 3$. Actually, Durkalec et al. (2017) estimate the satellite fractions at $z \sim 3$ to be high, $0.5 - 0.8$, which are inconsistent with the results of Ishikawa et al. (2016b). Durkalec et al. (2017) use the spectroscopically confirmed stellar-mass-threshold sample from the VIMOS Ultra Deep Survey, while the results of Ishikawa et al. (2016b) are based on the LBGs. Unknown bias in the sample selection could effect the satellite

 

fraction estimate.

The second is the evolution of the halo mass function. Since the number density of massive halos decreases more rapidly than that of less-massive halos with increasing redshift, the significant decrease of $f_{sat}$ could be explained by the decrease of the abundance of massive halos having many satellite galaxies. To examine this possibility, we calculate a subhalo fraction, $f_{sub}$, for stellar-mass-threshold samples, with a redshift-independent subhalo mass-host halo mass relation. In Behroozi et al. (2013b), the subhalo mass function is written as

$$\frac{d\phi}{dM_s} = \frac{C(a)}{M_s} \int_{M_s}^{\infty} M_c \frac{dn}{dM_c} d(\log M_c), \tag{32}$$

where $C(a)$ is proportionality factor (see Behroozi et al. 2013b) as a function of scale factor, and $\frac{dn}{dM_c} = \frac{dn}{dM}(M_c, z)$ is the host halo mass function. $M_c$ is the host (central) halo mass, while $M_s$ is the subhalo mass defined by the peak halo mass. By differentiating Equation (32) with respect to $M_c$, we obtain the number density of $M_s$ subhalos hosted by halos of $M_c$:

$$\frac{d\phi}{dM_c dM_s} = \begin{cases} \frac{C(a)}{M_s \ln 10} \frac{dn}{dM_c} & \text{if } M_s < M_c, \\ 0 & \text{otherwise.} \end{cases} \tag{33}$$

Thus the subhalo fraction of the sample with the stellar mass threshold $M_*^{th}$ is

$$f_{sub} = \frac{\int_{M_h^{th}}^{\infty} dM_c \int_{M_s^{th}}^{\infty} dM_s \frac{d\phi}{dM_c dM_s}}{\int_{M_h^{th}}^{\infty} dM_c \int_{M_s^{th}}^{\infty} dM_s \frac{d\phi}{dM_c dM_s} + \int_{M_h^{th}}^{\infty} dM_c \frac{dn}{dM_c}} \tag{34}$$

where $M_h^{th}$ ($M_s^{th}$) is the threshold mass of the host halo (subhalo) corresponding to $M_*^{th}$ in the host halo-cental galaxy (subhalo-satellite galaxy) relation. We calculate $M_h^{th}$ from $M_*^{th}$ using the SHMR in Figure 20. We parameterize a $M_s^{th} - M_h^{th}$ relation (subhalo-host halo relation) as follows to reproduce $f_{sat}$ at $z \sim 0$:

$$\log M_s^{th} = 2 \log M_h^{th} - 14.0, \tag{35}$$

The halo mass dependence is needed to explain the decreasing trend of $f_{sat}$ with increasing $M_*^{th}$. We plot the calculated subhalo fractions with Equation (35) at $z \sim 0$, 4, and 5 in Figure 23 with the dashed curves. The calculated subhalo fractions do not reproduce the observed satellite fractions at $z \sim 4 - 5$, indicating that the evolution of the halo mass function cannot explain the significant decrease of $f_{sat}$ from $z \sim 0 - 2$ to $z \sim 4 - 5$.

The third is the evolution of the subhalo-satellite galaxy connection. In Figure 23, we also plot the subhalo fractions with the dotted curves using the following assumption;

$$\log M_s^{th} = 2 \log M_h^{th} - 12.3. \tag{36}$$

We can reproduce the observed satellite fractions with this assumption at $z \sim 4 - 5$, indicating that the $M_s^{th} - M_h^{th}$ relation could change from Equation (35) at $z \sim 0 - 2$ to Equation (36) at $z \sim 4 - 5$. In other words, at fixed $M_h^{th}$, the mass of subhalo having a galaxy whose stellar mass is the same as one in the $M_h^{th}$ host halo evolves drastically, by a factor of $\sim 50$ from $z \sim 2$ to $z \sim 4$. Although we cannot eliminate this possibility, it is hard to understand that the subhalo-host halo relation changes very rapidly only in $\sim 1.8$ Gyr between $z \sim 2$ and $z \sim 4$.

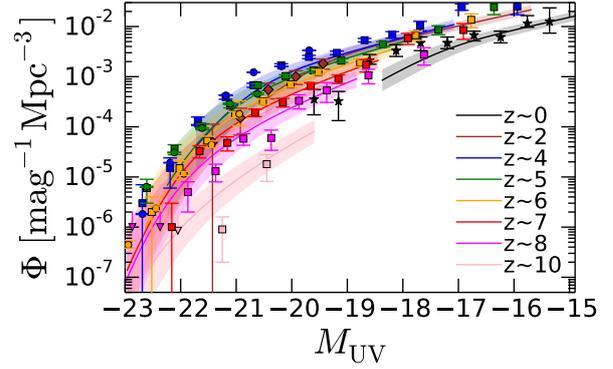

**Fig. 25.** Comparison of the UV luminosity functions. Solid curves denote the calculated luminosity functions from Equation (37). The points show the observed luminosity functions of Arnouts et al. (2005, stars), Oesch et al. (2010, diamonds), Ono et al. (2017, circles), and Bouwens et al. (2015, squares). The shades correspond to the 0.15 dex scatter in the $SFR/\dot{M}_h - M_h$ relation.

## 5.4 Fundamental $SFR/\dot{M}_h - M_h$ Relation

In Section 4.4, we find the fundamental $SFR/\dot{M}_h - M_h$ relation; the value of $SFR/\dot{M}_h$ at fixed $M_h$ does not significantly change beyond 0.15 dex at $z \sim 4 - 7$. Firstly, we examine whether this fundamental $SFR/\dot{M}_h - M_h$ relation is consistent with the observational results, i.e., the UV luminosity functions and cosmic SFRDs. We calculate the UV luminosity function at each redshift as follows:

$$\Phi(M_{UV}) = \frac{dn}{dM_h} \frac{dM_h}{dM_{UV}}. \tag{37}$$

From Equation (25), we can obtain the $M_{UV} - M_h$ relation and $\frac{dM_h}{dM_{UV}}$ at each redshift, since $\dot{M}_h$ can be expressed as a function of $M_h$ and $z$ (Behroozi et al. 2013b). Note that $M_{UV}$ is the observed absolute magnitude after extinction assuming the attenuation-UV slope ($\beta_{UV}$) relation (Meurer et al. 1999) and $\beta_{UV} - M_{UV}$ relations (Bouwens et al. 2014). We Consider 0.2 dex scatter in the halo mass ($\sigma_{\log M_h} = 0.2$), and use the satellite fractions in this study and literature. The calculated UV luminosity functions at $z \sim 0 - 10$ are plotted in Figure 25. We find that the calculated luminosity functions agree well with observed results given the 0.15 dex uncertainty in $SFR/\dot{M}_h$, indicating that our fundamental $SFR/\dot{M}_h - M_h$ relation is consistent with the observed UV luminosity functions. Tacchella et al. (2013) and Mason et al. (2015) conduct similar model calculations considering dust attenuation effects, and reproduce the UV luminosity functions at $z \sim 0 - 8$ and $z \sim 0 - 10$, respectively.

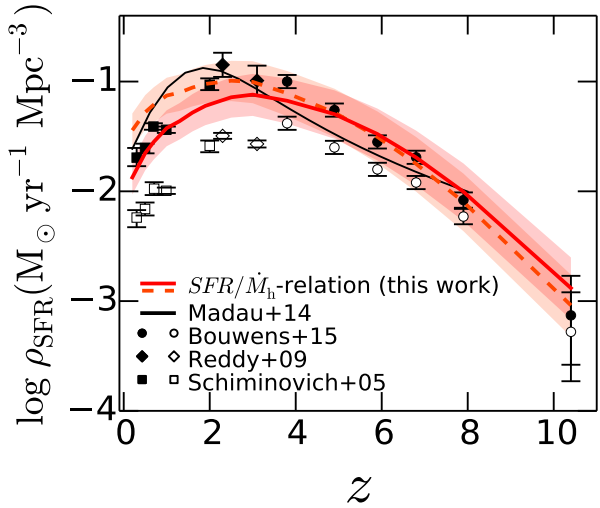

**Fig. 26.** Comparison of the cosmic SFRD. The red curves with the shade represent cosmic SFRDs derived from Equation (38). We assume Equation (25) and (26), which are constrained in this paper at $4 \lesssim z \lesssim 7$, for the solid and dashed curves, respectively. The filled circles, diamonds, and squares are cosmic SFRDs corrected for dust extinction from Bouwens et al. (2015), Reddy & Steidel (2009), and Schiminovich et al. (2005), respectively. Open symbols are densities before the dust corrections. The black solid curve is the fitting function from Madau & Dickinson (2014).

We also calculate the cosmic SFRDs as follows:

$$\rho_{\mathrm{SFR}} = \int dM_{\mathrm{h}} \frac{dn}{dM_{\mathrm{h}}} SFR(M_{\mathrm{h}}, z), \tag{38}$$

where $SFR(M_{\mathrm{h}}, z)$ is obtained from Equation (25). We compare our cosmic SFRDs with observations in Figure 26. We find that our calculation (the solid curve in Figure 26) well reproduces the overall observed trend of the cosmic SFRDs; the calculated densities increase from $z \sim 10$ to $4-2$, and decrease from $z \sim 4-2$ to 0. However, they could be underpredicted compared to observations at $z \sim 2$. Then, we use the evolving $SFR/\dot{M}_{\mathrm{h}} - M_{\mathrm{h}}$ relation, Equation (26), instead of Equation (25) to calculate $SFR(M_{\mathrm{h}}, z)$ in Equation (38). With Equation (26), our model-calculated cosmic SFRDs (the dashed curve in Figure 26) agree with the observations at the $\sim 1\sigma$ level at $z \sim 2$. These good agreements indicate that the evolution of the cosmic SFRDs is driven by the increase of the halo number density and the decrease of the accretion rate. From $z \sim 10$ to $4-2$, the number of halos at fixed masses increases (the upper panel in Figure 27), resulting the increase of the galaxy number density ($\Phi^*$ evolution in the UV luminosity function). From $z \sim 4-2$ to 0, the dark matter (and gas) accretion rate decreases (the middle panel in Figure 27), and the SFRs of galaxies at fixed dark matter halo masses decrease ($L^*$ evolution in the UV luminosity function). Because the cosmic SFRD is proportional to the galaxy number densities and SFRs, the calculated cosmic SFRD has a peak at $z \sim 2-4$ (the lower panel in Figure 27). These trends are consistent with the evolutions of the UV luminos-ity functions parameters (e.g., Bouwens et al. 2015; Parsa et al. 2016).

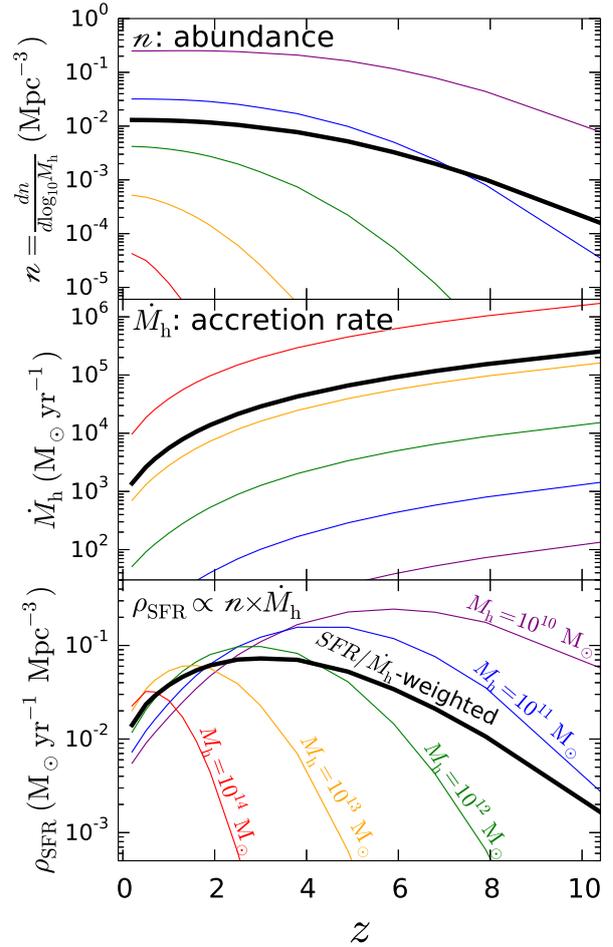

**Fig. 27.** Mechanism of the cosmic SFRD evolution. Upper panel: the purple, blue, green, orange, and red curves indicate the number density of halos of $M_{\mathrm{h}} = 10^{10}$, $10^{11}$, $10^{12}$, $10^{13}$, and $10^{14}$ $M_{\mathrm{h}}$, respectively. The black curve represents a weighted number density based on Equation (25). Middle panel: same as the upper panel but for the dark matter accretion rate. Lower panel: same as the upper panel but for the SFRD. Because the cosmic SFRD is proportional to the galaxy (halo) number densities and SFRs (accretion rates), the calculated cosmic SFRD has a peak at $z \sim 2-4$.

Secondly, we investigate whether the fundamental $SFR/\dot{M}_{\mathrm{h}} - M_{\mathrm{h}}$ relation is consistent with the observed SHMR redshift evolution. The stellar mass in the halo of $M_{\mathrm{h}}$ can be calculated as follows:

$$M_* = (1 - R) \int_{t=0}^{t=t(z)} dt \, SFR \tag{39}$$

$$= (1 - R) \int_{t=0}^{t=t(z)} dt \frac{SFR}{\dot{M}_{\mathrm{h}}} \dot{M}_{\mathrm{h}}, \tag{40}$$

where $R$ is the return fraction, or the mass fraction of each generation of stars that is put back into the inter-stellar and intergalactic medium by SNe and stellar winds. With the initial mass function, $\Phi(M)$, and the evolved mass of a star of initial mass $M$, $M_{\mathrm{evol}}(M)$, the return fraction is defined as follows:

$$R = \frac{\int dM \, (M - M_{\rm evol}(M)) \Phi(M)}{\int dM \, M \Phi(M)}. \tag{41}$$

If we assume that $SFR/\dot{M}_{\rm h}$ and $R$ do not depend on redshift in all the halo mass and redshift ranges, the stellar mass also does not depend on redshift,

$$M_* = (1 - R) \int_{t=0}^{t=t(z)} dt \frac{SFR}{\dot{M}_{\rm h}}(M_{\rm h}') \dot{M}_{\rm h} \tag{42}$$

$$= (1 - R) \int_{M_{\rm h}'=0}^{M_{\rm h}'=M_{\rm h}} dM_{\rm h}' \frac{SFR}{\dot{M}_{\rm h}}(M_{\rm h}'). \tag{43}$$

Thus the SHMR ($= M_*/M_{\rm h}$) should be independent on the redshift. This is not consistent with the observational result of the SHMR redshift evolution at the fixed halo mass. However, the return fraction is not constant over the star formation history. Behroozi et al. (2013b) calculate the return fraction (the rate of stellar mass loss in their paper) using the FSPS package (Conroy et al. 2010; Conroy & Gunn 2010) under the Chabrier (2003) IMF and Bruzual & Charlot (2003) stellar evolution tracks. From their fitting formula, we find that $R$ from a single stellar population varies from 0 to 0.36 with increasing the time since its formation from 0 to 1.7 Gyr, corresponding to the age of the universe at $z = 3.8$. Thus, the SHMR at $z \sim 4$ could be lower than that at $z \sim 7$ by a factor of $\sim 0.64$. In addition, it is not clear whether the ratio of $SFR/\dot{M}_{\rm h}$ is independent on the redshift in all the halo mass and redshift ranges, since the ranges we prove in this study are limited, $4 \lesssim z \lesssim 7$, and $4 \times 10^{10} \ M_\odot \lesssim M_{\rm h} \lesssim 4 \times 10^{12} \ M_\odot$. If $SFR/\dot{M}_{\rm h}$ at $4 < z < 7$ is lower than that at $z > 7$ in $M_{\rm h} < 4 \times 10^{10} \ M_\odot$, the SHMR at $z \sim 4$ could also be lower than that at $z \sim 7$. Theoretically speaking, the UV background radiation suppresses the star formation in low mass halos such as $M_{\rm h} \sim 10^7 - 10^8 \ M_\odot$ after cosmic reionization (Susa & Umemura 2004), and the suppression could be stronger in more clumpy galaxies (see also Susa & Umemura 2000). Shibuya et al. (2016) suggest a possible increase of the clumpy galaxy fraction from $z \sim 8$ to $z \sim 4$. Thus, $SFR/\dot{M}_{\rm h}$ at $4 < z < 7$ could be lower than that at $z > 7$ in low mass halos, where the UV background radiation suppresses the star formation. In summary, the fundamental $SFR/\dot{M}_{\rm h} - M_{\rm h}$ relation can be consistent with the SHMR evolution.

Finally, we discuss what makes the fundamental $SFR/\dot{M}_{\rm h} - M_{\rm h}$ relation, especially at $z \sim 4 - 7$. Since the dark matter accretion is associated with the gas accretion, this relation indicates that the SFR is determined by the gas accretion rate and the halo mass, but not redshift. In other words, a certain percentage of the accreted gas is always converted to stars at fixed halo mass. This percentage is not 100% like $SFR/\dot{M} = f_{\rm b} = 0.159$, but $\sim 3 - 10\%$ depending only on $M_{\rm h}$. We cannot explain this solely by the free fall timescale, feedback, or gas cooling, which are related to the gas number density, gravitational potential, or virial temperature, respectively, because these quantities vary with redshift at fixed halo mass.

Two possibilities are considered to explain this fundamental $SFR/\dot{M}_{\rm h} - M_{\rm h}$ relation. One is the combination of these effects. The free fall time scale and gravitational potential become shorter and deeper with higher redshift for a given halo mass, leading to higher SFR, while the higher virial temperature suppresses the star formation. Thus the redshift evolutions of these quantities could cancel each other, resulting the redshift-independent relation.

The other is a process depending only on the halo mass. If there is another star formation or feedback process whose efficiency is only a function of $M_{\rm h}$, we may explain the fundamental $SFR/\dot{M}_{\rm h} - M_{\rm h}$ relation. For example, if the efficiency of the AGN feedback depends only on the black hole mass and it is a function of the halo mass, such AGN feedback could make the high-mass part of the $SFR/\dot{M}_{\rm h} - M_{\rm h}$ relation. In both cases, the weak evolution suggests that the $SFR/\dot{M}_{\rm h} - M_{\rm h}$ relation is a fundamental relation in high-redshift ($z \gtrsim 4$) galaxy formation whose star-formation activities are regulated by the dark-matter mass assembly.

# 6 Summary

We obtain clustering measurements from 579,492 LBGs down to $i \sim 26$ mag at $z \sim 4 - 6$ identified in the $\sim 100 \ \deg^2$ HSC SSP survey data. Combining the HOD models and the previous clustering measurements of faint galaxies in Harikane et al. (2016), we investigate the non-linear halo bias effect, $M_{\rm UV} - M_{\rm h}$ relation, SHMRs, satellite fractions, and $SFR/\dot{M}_{\rm h}$. Our major findings are summarized below:

1. We clearly identify the non-linear halo bias effect in our ACFs. The HOD models with the linear scale-independent halo bias (the linear HOD model) can reproduce the ACFs at small ($\theta < 10''$) and large ($\theta > 90''$) scales, but underestimate in $10'' < \theta < 90''$, the transition scale between 1- and 2-halo terms. The HOD models with the non-linear scale-dependent halo bias (the non-linear HOD model) fill the gaps between the observed ACFs and predictions of the linear HOD model, but are still not enough in some subsamples.

2. We reveal the $M_{\rm UV} - M_{\rm h}$ relations at $z \sim 4 - 6$ over two orders of magnitude in $M_{\rm h}$. The halo mass of LBGs ranges from $4 \times 10^{10} \ M_\odot$ to $4 \times 10^{12} \ M_\odot$, and increases with brighter magnitude. We find the redshift evolution of the $M_{\rm UV} - M_{\rm h}$ relation at more than $5.2\sigma$ confidence level at $z \sim 4 - 7$; $M_{\rm h}$ becomes smaller with increasing redshift at fixed $M_{\rm UV}$.

3. We calculate the SHMRs at $z \sim 4 - 6$ for more massive halos than Harikane et al. (2016), resulting in typically $\sim 10^{-2}$ at $M_{\rm h} \sim 10^{12} \ M_\odot$. The SHMRs increase with increasing $M_{\rm h}$, but their slopes become flatter probably due to inefficient gas cooling and/or the AGN feedback in massive halos. We find that the SHMR increases from $z \sim 4$ to 7 by a factor of $\sim 4$ at $M_h \simeq 1 \times 10^{11} \, M_\odot$, while the SHMR shows no strong evolution in the similar redshift range at $M_h \simeq 1 \times 10^{12} \, M_\odot$.

4. We calculate the satellite fraction, $f_{sat}$, for our LBGs at $z \sim 4$ and 5. The satellite fractions are $f_{sat} = 3 \times 10^{-3} - 8 \times 10^{-2}$, and smaller in larger stellar mass threshold.

5. We find a tight relation of $SFR/\dot{M}_h - M_h$ showing no significant evolution beyond 0.15 dex in the wide-mass range over $z \sim 4 - 7$. This weak evolution suggests that the $SFR/\dot{M}_h - M_h$ relation is a fundamental relation in high-redshift ($z \gtrsim 4$) galaxy formation whose star formation activities are regulated by the dark matter mass assembly. Assuming this fundamental relation, we calculate UV luminosity functions and cosmic SFRDs (Madau-Lilly plot) over $z = 0 - 10$. We find that our calculation results explain the overall evolution of the UV luminosity function very well. Moreover, the cosmic SFRD evolution from our calculation has peaks at $z \sim 2 - 4$, agreeing with the one obtained by observations. This agreement suggests that the increase of the SFRD from $z \sim 10$ to $4 - 2$ is mainly driven by the increase of the halo abundance, and that the decrease of the SFRD from $z \sim 4 - 2$ to 0 is explained by the decrease of the dark matter (and gas) accretion rate.

## Acknowledgments

We thank the anonymous referee for a careful reading and valuable comments that improved the clarity of the paper. We are grateful to Shogo Ishikawa and Mimi Song for providing their data table. We thank Frank C. van den Bosch, Richard S. Ellis, Steven L. Finkelstein, Seiji Fujimoto, Kenji Hasegawa, Peter W. Hatfield, Akio K. Inoue, Charles Jose, Andrey V. Kravtsov, Kentaro Nagamine, Masayuki Tanaka, Masayuki Umemura, and Zheng Zheng for their useful comments and discussions.

The Hyper Suprime-Cam (HSC) collaboration includes the astronomical communities of Japan and Taiwan, and Princeton University. The HSC instrumentation and software were developed by the National Astronomical Observatory of Japan (NAOJ), the Kavli Institute for the Physics and Mathematics of the Universe (Kavli IPMU), the University of Tokyo, the High Energy Accelerator Research Organization (KEK), the Academia Sinica Institute for Astronomy and Astrophysics in Taiwan (ASIAA), and Princeton University. Funding was contributed by the FIRST program from Japanese Cabinet Office, the Ministry of Education, Culture, Sports, Science and Technology (MEXT), the Japan Society for the Promotion of Science (JSPS), Japan Science and Technology Agency (JST), the Toray Science Foundation, NAOJ, Kavli IPMU, KEK, ASIAA, and Princeton University.

This paper makes use of software developed for the Large Synoptic Survey Telescope. We thank the LSST Project for making their code available as free software at http://dm.lsst.org

The Pan-STARRS1 Surveys (PS1) have been made possible through contributions of the Institute for Astronomy, the University of Hawaii, the Pan-STARRS Project Office, the Max-Planck Society and its participating institutes, the Max Planck Institute for Astronomy, Heidelberg and the Max Planck Institute for Extraterrestrial Physics, Garching, The Johns Hopkins University, Durham University, the University of Edinburgh, Queen's University Belfast, the Harvard- Smithsonian Center for Astrophysics, the Las Cumbres Observatory Global Telescope Network Incorporated, the National Central University of Taiwan, the Space Telescope Science Institute, the National Aeronautics and Space Administration under Grant No. NNX08AR22G issued through the Planetary Science Division of the NASA Science Mission Directorate, the National Science Foundation under Grant No. AST-1238877, the University of Maryland, and Eotvos Lorand University (ELTE) and the Los Alamos National Laboratory.

This work is supported by World Premier International Research Center Initiative (WPI Initiative), MEXT, Japan, and KAKENHI (15H02064) Grant-in-Aid for Scientific Research (A) through Japan Society for the Promotion of Science (JSPS), and a grant from the Hayakawa Satio Fund awarded by the Astronomical Society of Japan. Y.H. acknowledges support from the Advanced Leading Graduate Course for Photon Science (ALPS) grant and the JSPS through the JSPS Research Fellowship for Young Scientists.